\documentclass[preprint,showpacs,preprintnumbers,amsmath,amssymb]{revtex4}
\usepackage{graphicx}% Include figure files
\usepackage{dcolumn}% Align table columns on decimal point
\usepackage{bm}% bold math
\usepackage{amssymb}
\usepackage{amsmath}
\usepackage{latexsym}

\begin{document}

\title{Close Binary Galaxies:  Application to Source of Energy and Expansion in   Universe}

\author{V.V. Sargsyan$^{1,2}$, H. Lenske$^{2}$, G.G. Adamian$^{1}$, and N.V. Antonenko$^{1}$}

\affiliation{
$^{1}$Joint Institute for Nuclear Research, 141980 Dubna, Russia,\\
$^{2}$Institut f\"ur Theoretische Physik der
Justus--Liebig--Universit\"at, D--35392 Giessen, Germany}

 \date{\today}

\begin{abstract}

Applying the microscopic nuclear physics ideas for fusion reactions to macroscopic  galactic
systems, we study the evolution of the compact   binary galaxy in mass asymmetry (transfer) coordinate.
The conditions for the formation of stable symmetric binary galaxy are analyzed.
The role of   symmetrization  of asymmetric binary galaxy in the transformation
of  gravitational energy into   internal energy of   galaxies accompanied by the release of a
large amount of energy during the symmetrization process is revealed.
%The relationship between the asymmetry and expansion of the universe is analyzed.
\end{abstract}


\maketitle

\section{Introduction}\label{sec:Intro}
%
%The study of the physical features and motions of a double star has proven very fruitful for the construction of stellar evolution.
%The special and ever-increasing interest in double galaxies is attracted by the problem of determining the mass of the galaxy.
%
The  observations of binary galaxies provide us the most direct and, therefore,
the most reliable data on the masses and evolution of galaxies \cite{Karachentsev,Karachentsev2,Pipino}.
The mass of a galaxy can be estimated with the virial theorem (the virial method), which is based on
the assumption of the statistical balance of the kinetic and potential energy
in the system. The mass of di-galaxies can be calculated using the
simple premise of closed Keplerian motions of galaxies as two point-like masses
(so-called the orbital motion method) \cite{Karachentsev}. This   orbital motion method seems to be more reliable
because of  lower uncertainties \cite{Karachentsev}.
%
%Comparing the orbital masses of a di-galaxy with the sum of individual masses determined
%from rotation curves, the width of the radio profile of 21 cm or the %dispersion of
%stellar velocities in the central regions of the components
%
The observations of binary galaxies show that the bulk of mass of
galaxies complies with  their standard optical (visible) boundaries.
Strong support for this approach is obtained
by the observation that
the di-galaxy systems  have a low ratio of the average orbital mass to luminosity,
that is inconsistent with the hypothesis of an invisible
(virial) mass \cite{Karachentsev}.
%
%Thus, pairs of galaxies look like a drop-down link from a common
%chain of structures of various scales with signs of virial excess mass.
%
%Interest in studying the structure and dynamics of galaxy systems continues unabated.
%

In studies
of binary galaxies one can test the concepts
of their formation and evolution.
The evolutionary path of the binary galaxy is not equivalent to that of isolated single galaxies.
The mutual tidal interaction favors    strong emission features in interacting galaxies.
Many evidences have been gathered that the star formation processes in binary galaxy  occur
more frequently than in   single galaxies or members of galactic associations.
In comparison with single galaxies, the binary ones contain a
higher percentage of objects of early structural types,
the Seyfert objects,   blue Markarian galaxies, and quasars \cite{Karachentsev}.
The galaxies in binary systems reveal mutual correlation by a number of integral features:
luminosity, linear diameter, structural type, mass to luminosity ratio, the value of its own spin.
%
%In the elliptical binary galaxies, the amorphous symmetric atmospheres are usually observed.
%It should be noted that the relative number of elliptical galaxies increases
%with decreasing distance between the components of the pair.
As follows from the observations, the orbits of  binary galaxies are close to circular, implying
large angular momenta.
%The best agreement with the observation data occurs when the eccentricity of the orbits is about 0.25.
% Because of the almost circular nature of the motion, the binary galaxies have large angular momenta.
 For example,  in the binary spiral (elliptical) galaxies,
the ratio between the value of  orbital angular momentum and the sum of  spins of its components
is about 1.3--2.5 (10) \cite{Karachentsev}.

%------------------------------------------------------

The compact or close
binaries, which  form  di-galaxy compounds
with the average distances between the galaxies  of the
same order as the sum of their radii, are of great interest for galactic evolution  \cite{Karachentsev,Karachentsev2,Pipino}.
Because mass transfer is  an important observable in close
binaries, it is   necessary to study the evolution of the
system in the mass asymmetry (transfer) coordinate
$\eta = (M_1-M_2)/(M_1+M_2)$ where $M_k$  $(k =1, 2)$  are the galactic
masses  at fixed total mass $M=M_1+M_2$ of the system \cite{IJMPE,IJMPE2}.
In our previous works \cite{IJMPE,IJMPE2}, we used
classical
Newtonian mechanics and studied the evolution of   close binary stars in the mass asymmetry (transfer) coordinate.
For close binary stars,
there are a lot of experimental and theoretical literatures about
the mass transfer between two constituents \cite{Kopal:1978,Shore:1994,Hild:2001,Boya:2002,Eggleton:2006,Cher:2013,Tutuk,Qian}.
%
%The limits for the
%formation and evolution of the di-star systems have been derived and analyzed.
In Refs. \cite{IJMPE,IJMPE2}, we applied theoretical methods which are  successfully   applied to the
corresponding processes in nuclear systems where the mass asymmetry plays an
important role as the collective coordinate governing fusion of two heavy nuclei
\cite{Adamian:2012,Adamian:2014}. Nuclear dynamics, of course, is quite different
from the gravitational interactions in di-stars. Nuclear reactions are
dominated by short-ranged strong interactions, to which minor
contributions of long-range (repulsive) Coulomb and centrifugal forces are superimposed.
However,  extending the methods and results from the femto-scale of
microscopic nuclear physics   to macroscopic binary stellar
systems, we obtained that  the driving potentials  for the di-star systems resembles in many aspects
the driving potentials for the microscopic dinuclear systems \cite{IJMPE,IJMPE2,Adamian:2012,Adamian:2014}.
In the present paper, we  explore the landscape of   total potential
energy of the close binary galaxy
as a function of   mass asymmetry coordinate, searching
specifically for the evolution paths in the mass asymmetry coordinate and
mass transfer source of the transformation of   gravitational  energy
to other types of energy in the universe.

\section{Theoretical Method}\label{sec:Theory}

The differential of total energy of di-galaxy system as a function of relative distance ${\bf R}$ of two galaxies,
conjugate canonical momentum ${\bf P}$, and mass asymmetry coordinate $\eta$ reads as
\begin{eqnarray}
dE({\bf R},{\bf P},\eta)=\frac{\partial E}{\partial t}dt+\frac{\partial E}{\partial {\bf R}}d{\bf R}+\frac{\partial E}{\partial {\bf P}}d{\bf P}+\frac{\partial E}{\partial \eta}d\eta.
\label{dEtot}
\end{eqnarray}
The kinetic energy in $\eta$ is assumed to be small and can be disregarded.
%As we consider the two stars as a closed system with $\partial E/ \partial t=0$, the conservation of the total energy leads to the following condition (and vice %versa):
%\begin{eqnarray}
%\frac{\partial E}{\partial \eta}\frac{d \eta}{d t}=0
%\label{dEeta}
%\end{eqnarray}
%which implies $\partial E/ \partial \eta=0$ as general solution.
In the center of mass system, the total energy of   di-galaxy system is a sum of radial and orbital part of kinetic energies  and the potential energy.
For the reasons indicated below, we attach the orbital kinetic energy part to the galaxy-galaxy interaction $V$.
In this case,
% the expression of the total energy of the di-galaxy system reads as
\begin{eqnarray}
E=\frac{P_{R}^2}{2 \mu}+U,
\label{Etot}
\end{eqnarray}
where $P_{R}$ is the radial component of   momentum ${\bf P}$ and
$\mu=\mu(\eta)=\frac{M_1M_2}{M}=\frac{M}{4}(1-\eta^2)$ is the reduced mass.

The total potential energy
\begin{eqnarray}
U=U_1+U_2+V
\label{eq_pot}
\end{eqnarray}
of the di-galaxy system
is given by the sum of the potential energies $U_k$ ($k=1,2$) of   two galaxies and galaxy-galaxy interaction potential $V$.
The radiation energy is neglected because
 the absolute values of the gravitational energy and the intrinsic kinetic energy
 are much  larger.
The energy of the galaxy "$k$" is
\begin{eqnarray}
U_{k}=-\frac{G M_k^2}{2R_k},
\label{eq_pot2}
\end{eqnarray}
where $G$, $M_k$, and $R_k$ are the gravitational constant, mass, and radius of the galaxy, respectively.
The observational data  result in the relationship
$$R_k=\frac{1}{g}M_k^{n}$$
between radius and mass of the galaxy, where the value of constant $n$
is  in the interval $\left[\frac{2}{5},\frac{2}{3}\right]$ \cite{Karachentsev}.
Employing this expression,
we obtain
\begin{eqnarray}
U_{k}=-\frac{G g M_{k}^{2-n}}{2}.
\label{eq_pot1}
\end{eqnarray}
The value of the constant
$$g=\frac{M^n}{\left(R_{1i}^{1/n}+R_{2i}^{1/n}\right)^n}$$
is determined by the observed  total mass $M=M_1+M_2=M_{1i}+M_{2i}$  and
radii $R_{1i,2i}=\frac{1}{g}M_{1i,2i}^{n}$ of two constituents.

Because the two galaxies rotate with respect to each other around the common center of
mass, the galaxy-galaxy interaction potential
\begin{eqnarray}
V(R)=V_{\rm grav}(R) + V_{\rm rot}(R)=-\frac{GM_1M_2}{2R_m}
\label{eq_pot5nn}
\end{eqnarray}
contains, together with the gravitational energy $V_{\rm grav}(R)=-GM_1M_2/R$
of the interaction of two stars, the energy $V_{\rm rot}(R)=\mu v^2/2$ of orbital rotation.
In Eq. (\ref{eq_pot5nn}), $v(R)=(GM[2/R-1/R_m])^{1/2}$ and $R_m$ are the speed and
the semi-major axis of  the elliptical relative orbit, respectively.
For the di-galaxies considered, $v(R_m)\ll c$, where $c$ is the velocity of light, and one can
neglect the relativistic effects. Since  $GM_k/R_m\ll c^2$, the gravitational field  can be considered weak.
Because of these facts, we use the Newtonian law of gravity.
Using the Kepler's laws, we obtain
\begin{eqnarray}
R_m=\left(\frac{\mu_i}{\mu}\right)^2R_{mi},
\label{eq_pot5nnD}
\end{eqnarray}
where the index "{\rm i}" denotes the observed  reduced
mass $\mu_i=\mu(\eta_i)=\frac{M_{1i}M_{2i}}{M}=\frac{M}{4}(1-\eta_i^2)$
and distance $R_{mi}$ between the  galaxies of the initial binary system with mass asymmetry $\eta=\eta_i$.
Equation (\ref{eq_pot5nn}) is rewritten as
\begin{eqnarray}
V(R_m)=-\omega_V\frac{GM_1^3M_2^3}{2},
\label{eq_pot5nnn}
\end{eqnarray}
where
$$\omega_V=\left(M^2\mu_i^2R_{mi}\right)^{-1}.$$
The final expression for the total potential energy (\ref{eq_pot}) of the di-galaxy system is
\begin{eqnarray}
U=-\frac{G}{2}\left(g[M_1^{2-n} + M_2^{2-n}]+\omega_VM_1^3M_2^{3}\right).
\label{eq_pot6}
\end{eqnarray}
Expressing the masses in terms of the mass asymmetry coordinate $\eta$,  $M_1=\frac{M}{2}(1+\eta)$ and $M_2=\frac{M}{2}(1-\eta)$,
we rewrite   Eq. (\ref{eq_pot6}):
\begin{eqnarray}
U=-\alpha[(1+\eta)^{2-n}+(1-\eta)^{2-n}]-\beta[1-\eta^2]^{3},
\label{eq_pot7}
\end{eqnarray}
where
%$$\alpha=\omega_0\left(\frac{M}{2}\right)^{13/12}=1.644\frac{M_{\odot}^{2}}{R_{\odot}}\left(\frac{M}{2M_{\odot}}\right)^{13/12}$$
$$\alpha=\frac{Gg}{2}\left(\frac{M}{2}\right)^{2-n}$$
and
$$\beta=\frac{GM^2}{128}\frac{(R_{1i}^{1/n}+R_{2i}^{1/n})^4}{R_{mi}R_{1i}^{2/n}R_{2i}^{2/n}}.$$
To obtain $\alpha$ and $\beta$, we use the  observed values of $M$, $R_{1i}$, $R_{2i}$, and $R_{mi}=\frac{\pi}{4}X$, where
$X$ is projection of the linear distance between the components of  binary galaxy,
 from the catalogue of isolated pairs of galaxies \cite{Karachentsev}.

Employing Eq. (\ref{eq_pot7}), we  study the evolution of the di-galaxy system in the mass asymmetry coordinate $\eta$.
The extremal points of the potential energy as a function of $\eta$ are found by solving numerically the equation
\begin{eqnarray}
\frac{\partial U}{\partial\eta}=-\alpha[2-n][(1+\eta)^{1-n}-(1-\eta)^{1-n}]+6\beta\eta[1-\eta^2]^{2}=0,
\label{eq_pot9}
\end{eqnarray}
and the solutions are the fixed points of the evolution equation for the mass asymmetry.
As seen, Eq. (\ref{eq_pot9}) is solved for $\eta=\eta_m=0$.
At this value the potential  has an extremum which is a minimum if
$$\alpha  < \frac{3}{(1-n)(2-n)}\beta$$
or
$$R_{mi}<\frac{3\left(R_{1i}^{1/n}+R_{2i}^{1/n}\right)^{4+n}}{2^{4+n}(1-n)(2-n)R_{1i}^{2/n}R_{2i}^{2/n}}$$
and a maximum if
$$\alpha  > \frac{3}{(1-n)(2-n)}\beta.$$
The transition point is
$$\alpha_{cr}=\alpha=\frac{3}{(1-n)(2-n)}\beta.$$
If there is a minimum at $\eta=0$ ($\alpha<\alpha_{cr}$),
it is engulfed symmetrically by two barriers at $\eta=\pm\eta_b$.
As seen, the extremal points depend only on $R_{mi}$, $R_{1i,2i}$,  and $n$.
Note that for the touching binary system ($R_{mi}\approx R_{1i}+R_{2i}$), a minimum at $\eta=0$ exists because
the condition $\alpha  < \frac{3}{(1-n)(2-n)}\beta$ or
$\frac{3\left(R_{1i}^{1/n}+R_{2i}^{1/n}\right)^{4+n}}{2(1-n)(2-n)\left(4R_{1i}^{1/n}R_{2i}^{1/n}\right)^\frac{4+n}{2}}>1$
is  always satisfied since $\frac{3}{2(1-n)(2-n)}>1$ and
$\frac{\left(R_{1i}^{1/n}+R_{2i}^{1/n}\right)^{4+n}}{\left(4R_{1i}^{1/n}R_{2i}^{1/n}\right)^\frac{4+n}{2}}>1$.

Expanding Eq. (\ref{eq_pot9}) up to the third order in $\eta$ and solving it,
we obtain the position of these barriers at $\eta=\pm\eta_b$, where
\begin{eqnarray}
\eta_b&=&2^{-1/2}\left[\frac{36\beta-12(1-n)(2-n)\alpha}{36\beta+n(1-n^2)(2-n)\alpha}\right]^{1/2}\approx
2^{-1/2}\left[1-\frac{(1-n)(2-n)}{3}\frac{\alpha}{\beta}\right]^{1/2}\nonumber\\
&=&2^{-1/2}\left[1-\frac{2^{4+n}(1-n)(2-n)R_{mi}R_{1i}^{2/n}R_{2i}^{2/n}}{3\left(R_{1i}^{1/n}+R_{2i}^{1/n}\right)^{4+n}}\right]^{1/2}.
\label{eq etabb}
\end{eqnarray}
So, at $\alpha<\alpha_{cr}$ the potential energy as a function of $\eta$
has two symmetric maxima at $\eta=\pm\eta_b$ and the minimum at
$\eta=\eta_m=0$.
As seen from Eq. (\ref{eq etabb}),
$\eta_b < 2^{-1/2}\approx 0.71$.
%  if $\alpha\ll\frac{3}{(1-m)(2-m)}\beta$ or $R_{mi}\ll\frac{3(R_{1i}^{1/m}+R_{2i}^{1/m})^{4+m}}{2^{4+m}(1-m)(2-m)R_{1i}^{2/m}R_{2i}^{2/m}}$.
This   condition
 means that in the asymmetric binary system
 with   mass ratio   $M_{1i}/M_{2i} > (1+2^{1/2})^2\approx 6$
 the galaxies  move away from each other.
 Thus, di-galaxies  with $|\eta| > \eta_b$ are the unstable with respect to the separation and
unlikely to exist as a correlated, bound pair for sufficiently long time.
Indeed, the binary galaxies with a large mass
ratio are very rare objects \cite{Karachentsev}.

%Because
%\begin{eqnarray}
%V(R_m)=-\frac{G^2M^5}{128L^2}(1-\eta^2)^3<0,
%\label{eq_pot100}
%\end{eqnarray}
%the solutions of the equation
%\begin{eqnarray}
%\frac{\partial V(R_m)}{\partial \eta}{\large|_{\eta=\eta_m}}=\frac{3G^2M^5\eta_m(1-\eta_m^2)^2}{64L^2}=0
%\label{eq_pot10}
%\end{eqnarray}
%are $\eta_m=0$ and $\eta_m=\pm 1$ and
%$$\frac{\partial^2 V(R_m)}{\partial \eta^2}|_{\eta=\eta_m=0}=\frac{3G^2M^5}{64L^2}>0,$$
%$$\frac{\partial^2 V(R_m)}{\partial \eta^2}|_{\eta=\eta_m=\pm 1}=0,$$
%the star-star interaction potential $V(R)$ has a minimum at $R=R_m$ and $\eta=\eta_m=0$.
%Hence, an initially asymmetric binary system ( $|\eta|=|\eta_i|<\eta_b$) is
%driven to mass symmetry, implying a flow of mass towards equilibrium.

\section{Application to Close Binaries}\label{sec:Results}

In the calculations, we use  the observational data for the  galactic  linear diameters  $A_{25}(1i,2i)=2R_{1i,2i}$,
 the projection of the linear distance $X$ between the components of the binary galaxy, and the
 total orbital mass $M$ of the pair  from the catalogue of the isolated pairs of galaxies \cite{Karachentsev}.
 For the  close binary galaxies considered, the average relative distance $R_{mi}=\frac{4}{\pi}X$
 between galaxies is comparable with the sum  of radii of galaxies \cite{Karachentsev}.
 The mass transfer between galaxies in a   binary system is
 closely related to their radii and the relative distance.
 From the observational data for galaxies with large mass $M_{1i,2i}>10^{10}M_{\odot}$,
 where $M_{\odot}$ is the mass of the Sun,
 the dependence between the mass and  radius of the galaxy
 is extracted as $R_k=\frac{1}{g}M_k^{n}$, where  $n=2/5$
 and $g= M^{2/5}/\left(R_{1i}^{5/2}+R_{2i}^{5/2}\right)^{2/5}$ \cite{Karachentsev}.
 The masses of components of the binary system
 are defined as  $M_{1i,2i}=MR^{1/n}_{1i,2i}/\left(R^{1/n}_{1i}+R^{1/n}_{2i}\right)$.
  Note that the
    initial mass asymmetry $\eta=\eta_i$ depends on $n$:
 $\eta_i=\left(R^{1/n}_{1i}-R^{1/n}_{2i}\right)/\left(R^{1/n}_{1i}+R^{1/n}_{2i}\right)$.
 The elliptic, spiral, and
 mixed elliptic-spiral binary galaxies are considered below.

Various  di-galaxies have different $R_{mi}$ and $R_{1i,2i}$, and, correspondingly, the potential energy shapes are varying.
The  potential energies (driving potentials) $U(\eta)$ of the close elliptic, spiral, and elliptic-spiral di-galaxy systems
versus $\eta$  are presented in Figs.~1-5.
For all systems shown, except the binary galaxy 272 \cite{Karachentsev},
$\alpha< \alpha_{cr}=\frac{25}{8}\beta$ and,
thus, the potential energies have
symmetric barriers at $\eta=\pm\eta_b$ and the minimum at $\eta=\eta_m=0$.
The barrier in $\eta$ appears from the interplay
between the total gravitational energy $U_1+U_2$ of the galaxies
and the galaxy-galaxy interaction potential $V$.
These energies behave differently as a function of mass asymmetry:
$U_1+U_2$ decreases and $V$ increases
with changing $\eta$ from $\eta=0$ to $\eta=\pm 1$.
One should stress that the driving potentials $U(\eta)$ for the di-galaxy systems look
like the driving potentials for the microscopic dinuclear systems \cite{Adamian:2012,Adamian:2014}.
For the close binary star  systems, the same conclusion was drawn in Refs. \cite{IJMPE,IJMPE2}.

An important question is what extent the results depend on the choice of parameters.
In  Fig. 2, we investigate the effect of varying the mass-radius power-law relation by using $n=2/3$.
As seen by comparison to Fig. 1, obtained with $n=2/5$,
the shapes of the potentials in both calculations are qualitatively the same,
except that the  ratio $U(\eta=\eta_b)/U(\eta=0)$ increases with   the value of $n$.
The central result of these investigations is that
all binary systems evolve to the symmetric
configurations with $\eta=0$.

The evolution path of  close binary galaxy   depends on the initial  mass asymmetry $\eta=\eta_i$ at its formation.
If the original di-galaxy is asymmetric, but $|\eta_i|<\eta_b$, then it is energetically favorable to evolve
in  $\eta$ to a configuration in the global minimum at $\eta=0$, that is, to form a symmetric  di-galaxy system.
%This means that the substance is transferred from the heavy component to the light component.
The matter of  the heavy partner can move freely to an adjacent light galaxy
enforcing the symmetrization of di-galaxy without additional driving energy.
%If   $|\eta_i|<\eta_b$ and $\eta_i\ne 0$,  the di-star system is unstable
%and evolves to the symmetric di-star system with $\eta=0$, thus
%enforcing the symmetrization of di-star.
%This means that matter is transferred from the heavy star  to the light star even without additional external energy.
The symmetrization  of an initially asymmetric binary galaxy leads to the decrease of potential energy $U$,
thus transforming of the potential energy into internal kinetic energy.
%The resulting symmetric di-star is created at large excitation energy.
For example, for the close elliptic binary galaxies
194 ($\eta_i$=0.28),
279 ($\eta_i$=0.50),
and
554 ($\eta_i$=0.25) \cite{Karachentsev},
the internal energies of galaxies
will
increase during symmetrization
by the amount
$\Delta U=U(\eta_i)-U(\eta=0)=$
$2\times 10^{50}$,
$3\times 10^{51}$,
and
$2\times 10^{50}$ J,
respectively (see Fig. 1 and Table I).
Because   most of close binary galaxies are asymmetric ones,
the symmetrization process leads to the release of a
large amount of energy in these systems
%, thus being an important source of energy in the universe
(see Tables I--III).

If   $|\eta_i|>\eta_b$ or $\eta_b=0$,  the di-galaxy system is
unstable and evolves towards more asymmetric system, thus,
enforcing the asymmetrization of the di-galaxy
(one galaxy "swallows" the other galaxy).
The matter is flowing freely from the light galaxy to the
heavy one. That process is sustained  without implementing additional external energy.
This mode of evolution also leads to  energy release.
Representative examples for this mode of evolution are the
%in Ref. \cite{Karachentsev} only
three cases of close spiral binary galaxies, namely the galaxies
206 ($|\eta_i|=0.45$, $\eta_b=0.34$), 243 ($|\eta_i|=0.31$, $\eta_b=0.07$),
and 439 ($|\eta_i|=0.63$, $\eta_b=0.56$) \cite{Karachentsev}, respectively,
for which  $|\eta_i|>\eta_b$ (Fig. 5).

Mergers of binary galaxy with $|\eta_i|<\eta_b$
can occur only by overcoming the barrier   at  $\eta=+\eta_b $ or $\eta=-\eta_b $.
With decreasing ratio $\alpha/\beta$,  the value of this barrier $B_\eta=U(\eta_b)-U(\eta_i)$ increases.
Because the barriers $B_{\eta}$ in $\eta$ are quite large for the  systems with $|\eta_i|<\eta_b$
 in Tables I--III,
the asymmetrization of the   di-star system by the thermal diffusion in mass asymmetry coordinate
is strongly suppressed.
Most likely,  the  barrier in $\eta$ is the reason why very asymmetric close binary galaxies
with   $|\eta_i|>\eta_b$   are rarely observed.
This imposes restrictions on the asymmetric configurations with $|\eta|>\eta_b$
of the di-galaxy systems. There can not be a stable di-galactical system
with a very light galaxy, with only a fraction of  the mass of the heavy partner.

If the value of  $\alpha$ becomes larger than  $\alpha_{cr}$, the minimum in $U(\eta)$ disappears,
the inverted $U$-shaped potential shape with maximum at $\eta=0$ forms,
and the di-galaxy asymmetrization drives the system finally apart towards separation. Hence, with increasing mass asymmetry, two constituents of
the binary galaxy move away from each other and finally separate into single galaxies.
The spiral binary galaxy 272 (see Fig. 4 and Table III) is a good candidate  for such evolution.

Our calculations show that  the    double galaxy will undergo either
symmetrization or asymmetrization depending critically on the mass asymmetry.
%, with the second process being rarer than the first.
Asymmetrization may start as a merger process, but once the critical mass asymmetry is reached, the system
is driven apart and ends up by two isolated galaxies of a changed mass ratio.
Hence, this scenario corresponds to an incomplete merger.
%
%but there can be no complete merging due to the separation of
%the components from each other with increasing asymmetry.
%Fragmentation processes and consequently a partial merger are  the more likely scenarios.

So, the source of expansion of a binary galaxy is the transfer of mass from the lighter component to the heavy one.
A necessary and sufficient condition for this is the fulfillment of the inequality
$$R_{mi}>\frac{0.15\left(R_{1i}^{5/2}+R_{2i}^{5/2}\right)^{22/5}}{R_{1i}^{5}R_{2i}^{5}}.$$
This mechanism presented can be generalized for multiple galaxies, groups of galaxies and galaxy associations.
%The separation of components from each other can be represented
%as an analogue of the expansion of the universe within
%the framework of a binary system.

%--------------------

\section{Summary}\label{sec:Summary}

We have shown that the mass asymmetry (transfer) collective degree of freedom plays
a comparable important role   in macroscopic object as well
as in microscopic dinuclear systems. In close binary galaxy,
the   mass asymmetry coordinate
can govern the asymmetrization (the transfer of mass from the lighter component to the heavy one)
and symmetrization (the transfer of mass from the heavier component to the light one)
  of the system.
The symmetrization of binary galaxies leads to the release of a
large amount of energy about $10^{48-52}$~J, thus reaching the energy release of novae or even close to supernovae events.
Thus, the symmetrization of close binary galaxy  due to the mass transfer is one of the important
sources of the transformation of the gravitational  energy  to other types of energy, like radiation energy, in the universe.
The symmetrization of binary system will lead to $M_1/M_2\to 1$,
  $L_1/L_2\to 1$ ($L_i$ are the luminosities of   galaxies), $R_1/R_2\to 1$
which are observable quantities.
Asymmetrization is equivalent to the incomplete merging of components of a binary system.
The asymmetrization is also the source of expansion of a binary galaxy.
The separation of components from each other can be represented
as an analogue of the expansion of the universe within
the framework of a binary system.
The conditions under which either asymmetrization or symmetrization processes
are realized have been determined and investigated.
These conditions depend mainly on the relative distance between the galaxies and their linear sizes.
The limitation of the binary system existence at the mass ratios    $M_{1i}/M_{2i} > 6$
 has been derived.

In the frame of our model, we can also perform dynamic calculations of the evolution of binary system
in the mass asymmetry coordinate. For example, we can calculate the relaxation (symmetrization) time and the
asymmetrization time. In addition to the total potential energy, it is also necessary to calculate the mass parameter
for this coordinate. But this extension of our model is beyond the scope of the present paper.

\section{Acknowledgements}
V.V.S.   acknowledge  the partial
supports from the Alexander von Humboldt-Stiftung (Bonn).
This work was partially supported by  Russian Foundation for Basic Research (Moscow)  and
DFG (Bonn), contract Le439/16.

%\newpage

\begin{figure} [ht]
\centering
{\includegraphics[width=0.49\linewidth]{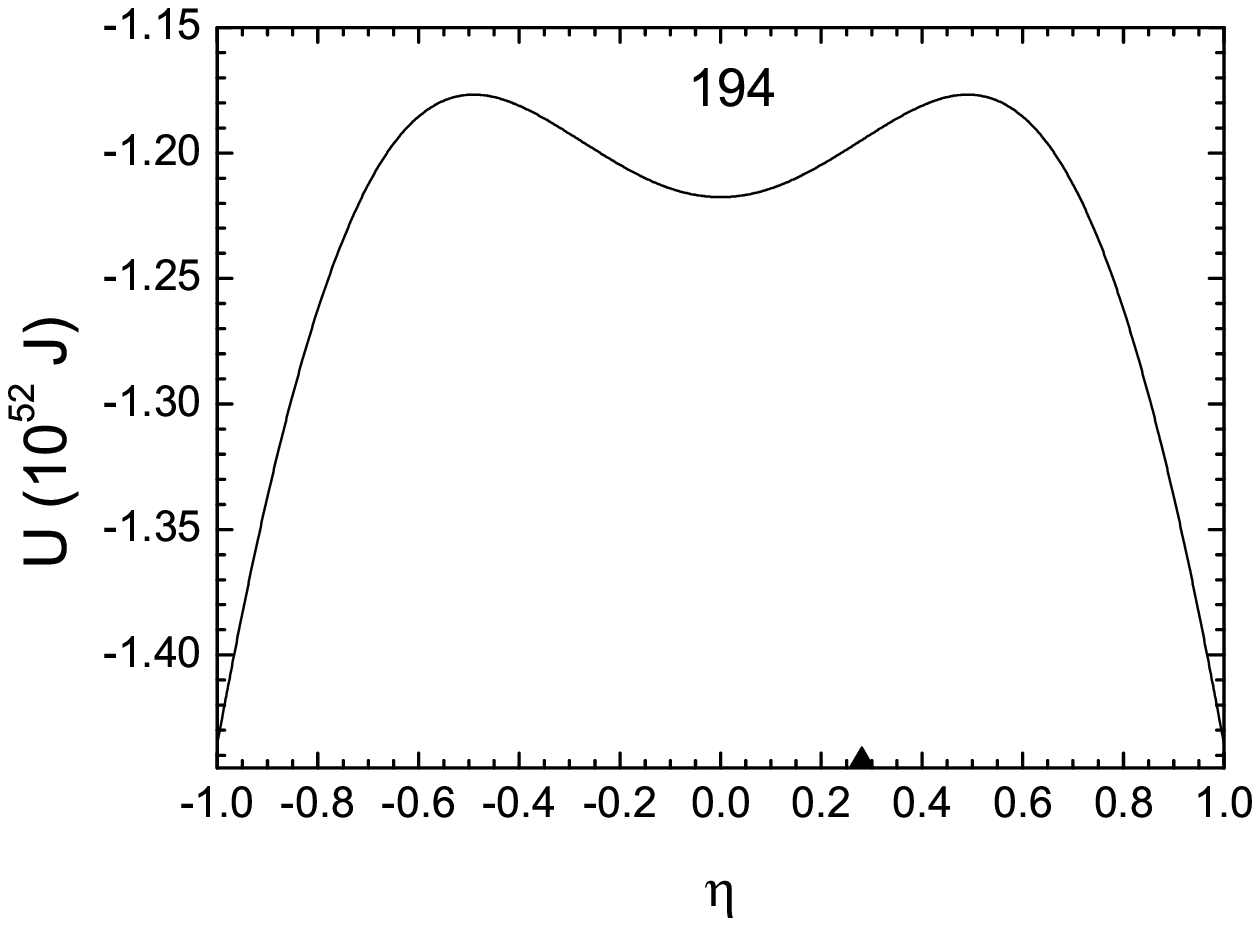}
\includegraphics[width=0.49\linewidth]{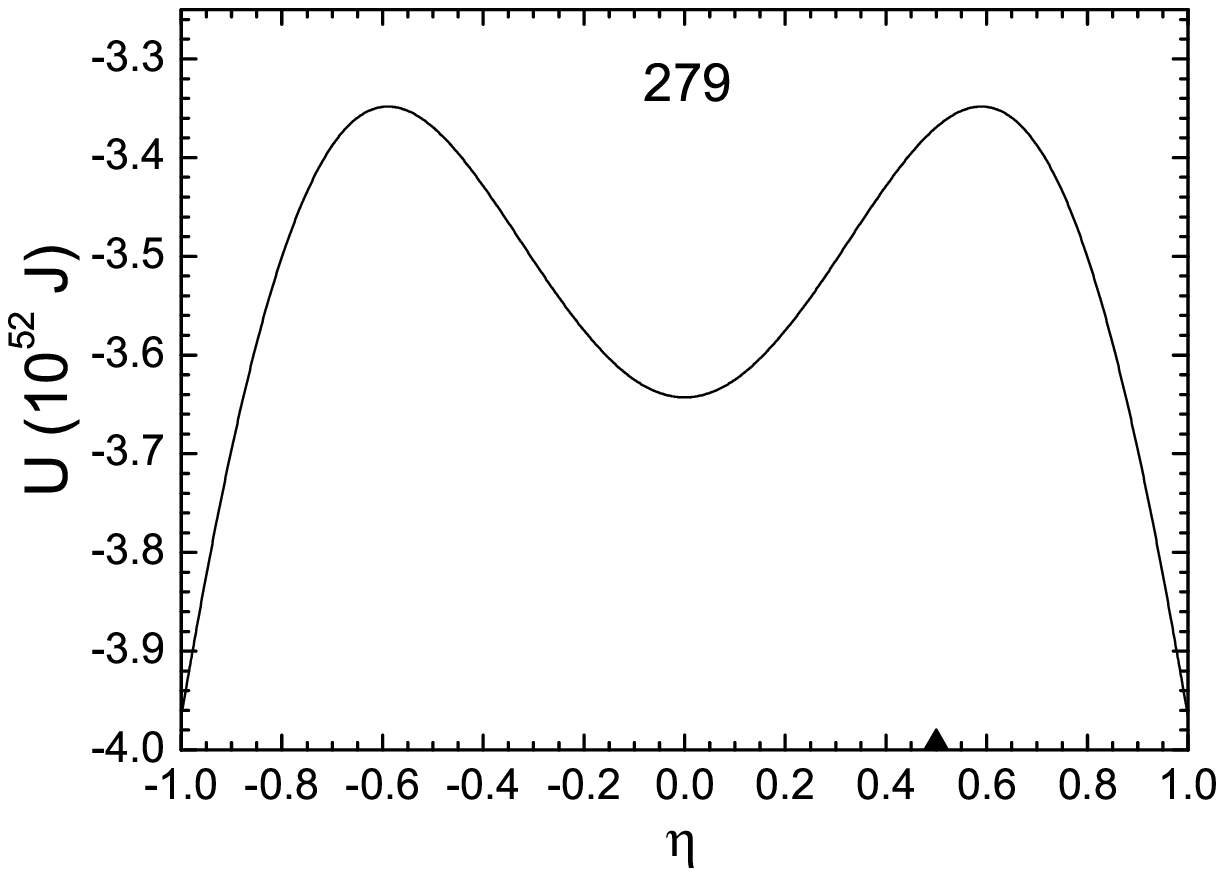}}
{\includegraphics[width=0.49\linewidth]{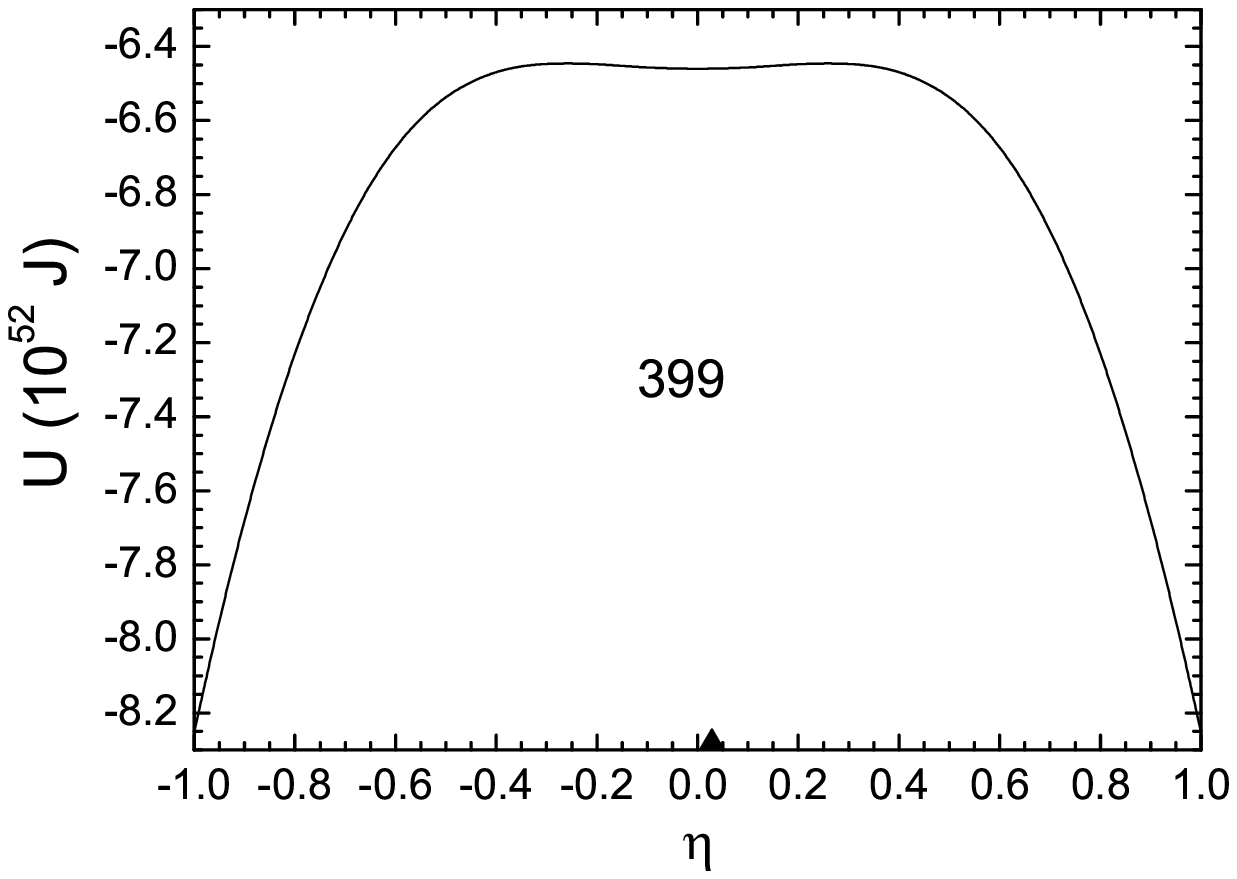}
\includegraphics[width=0.49\linewidth]{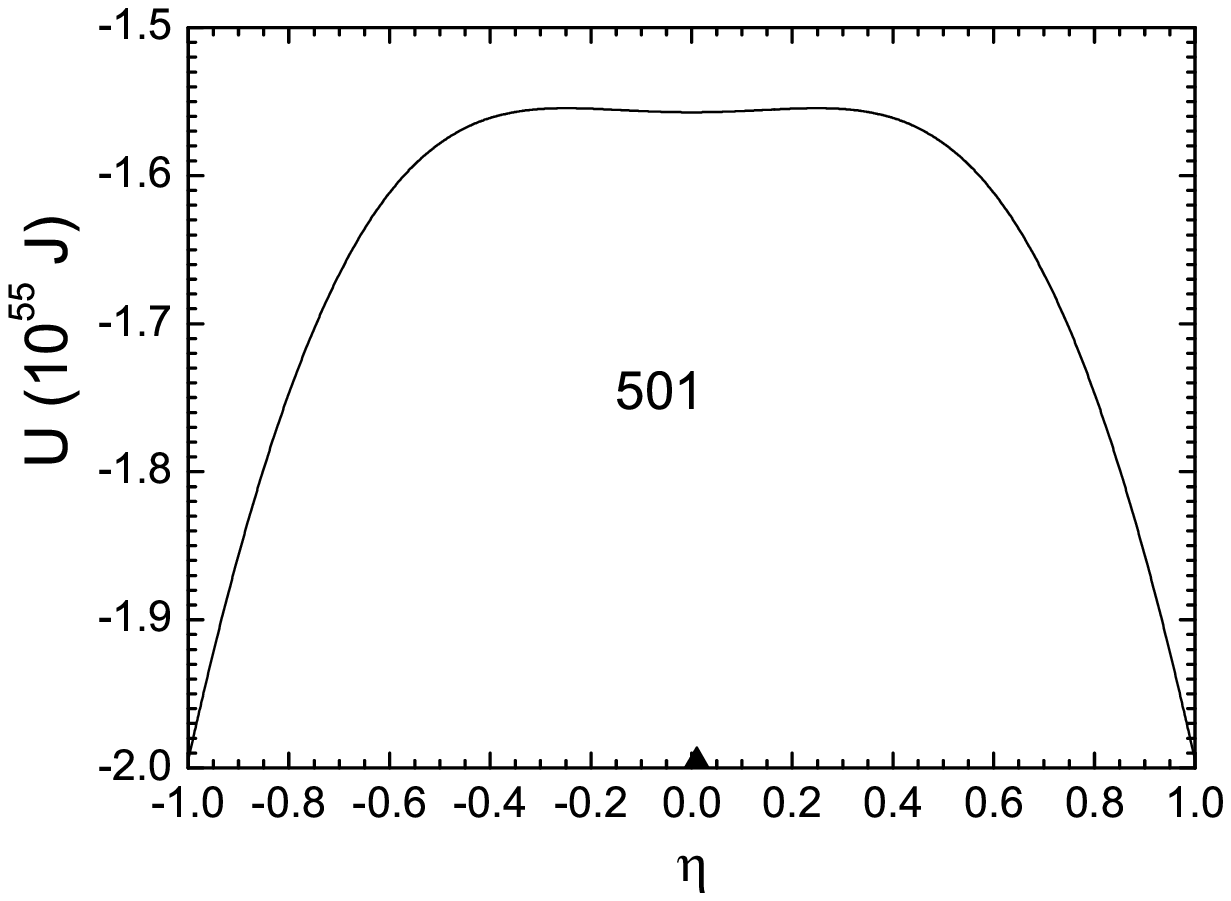}}
{\includegraphics[width=0.49\linewidth]{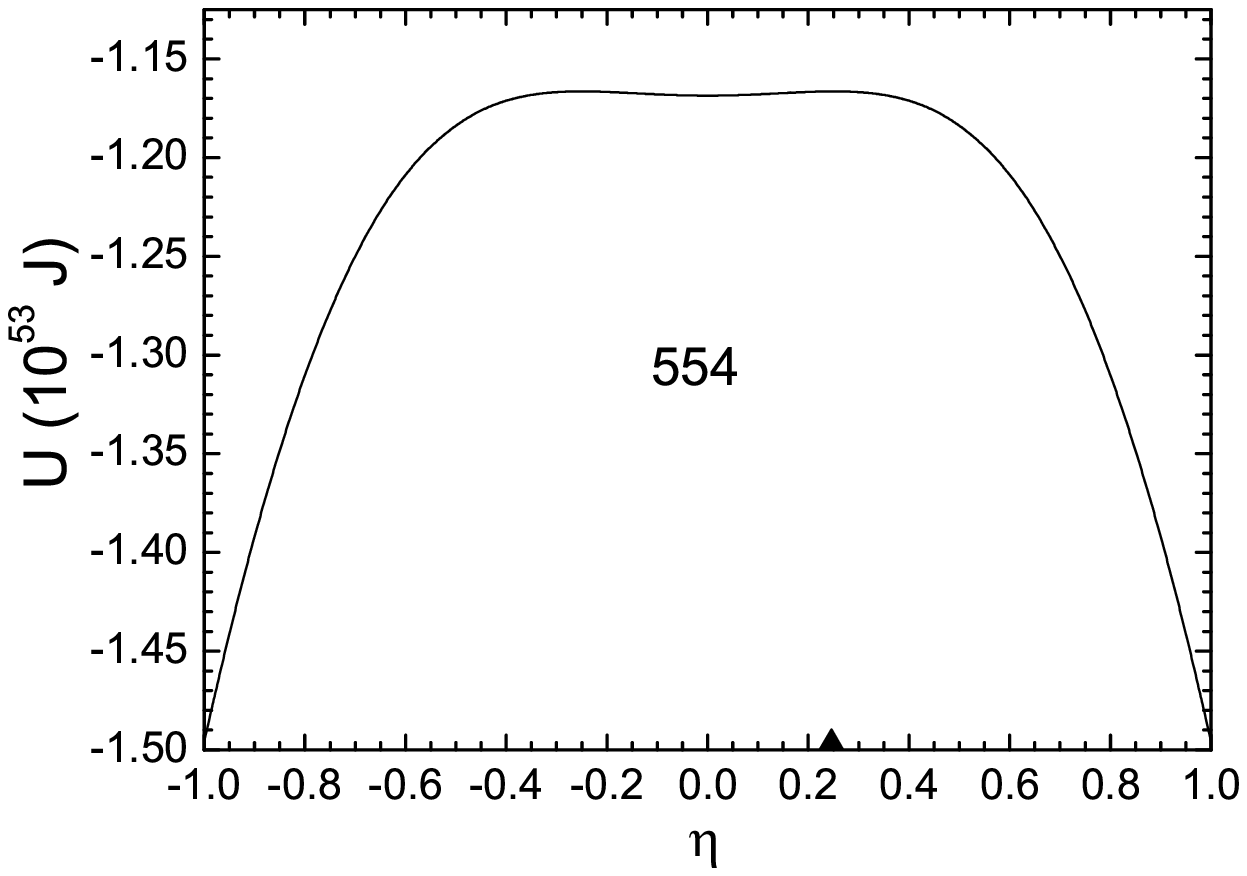}
\includegraphics[width=0.49\linewidth]{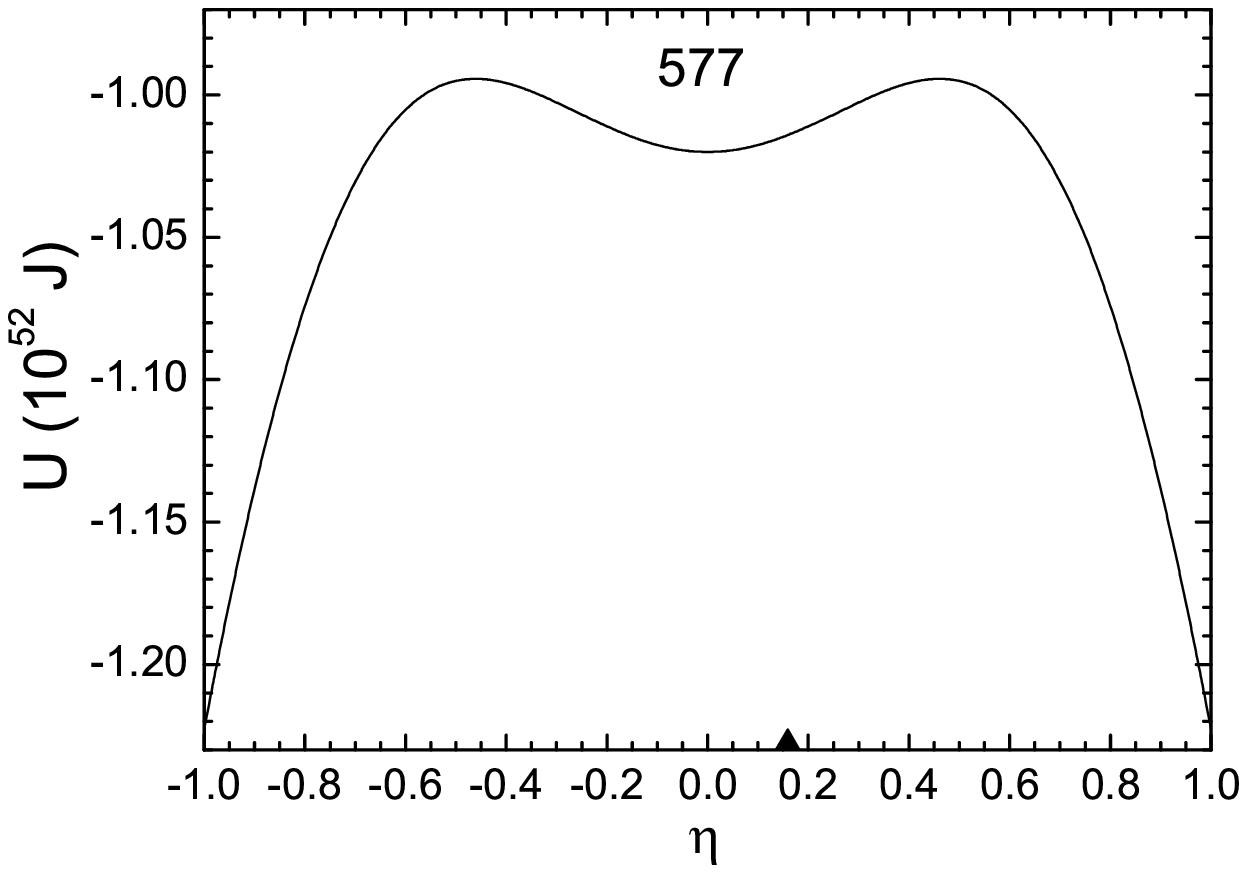}}
\caption{The calculated total potential energies $U$
vs $\eta$ for the indicated (numbers) close  binary elliptic galaxies at $n=2/5$.
The numbers are serial numbers of binary galaxies
in the catalogue of isolated pairs of galaxies \cite{Karachentsev}.
The arrows on $x$-axis show the corresponding
initial $\eta_i$ for binary galaxies.
}
\label{1_fig}
\end{figure}

\begin{figure} [ht]
\centering
{\includegraphics[width=0.49\linewidth]{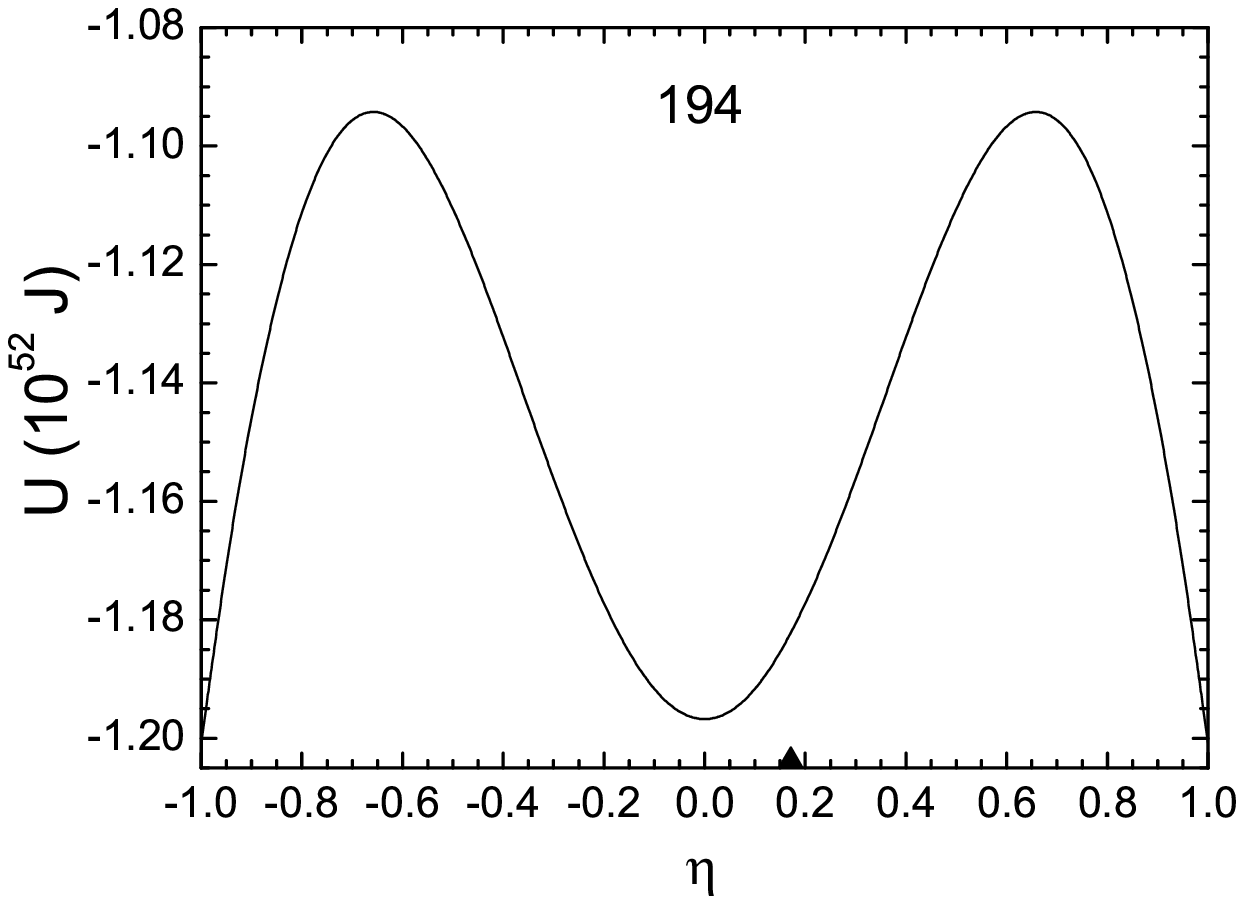}
\includegraphics[width=0.49\linewidth]{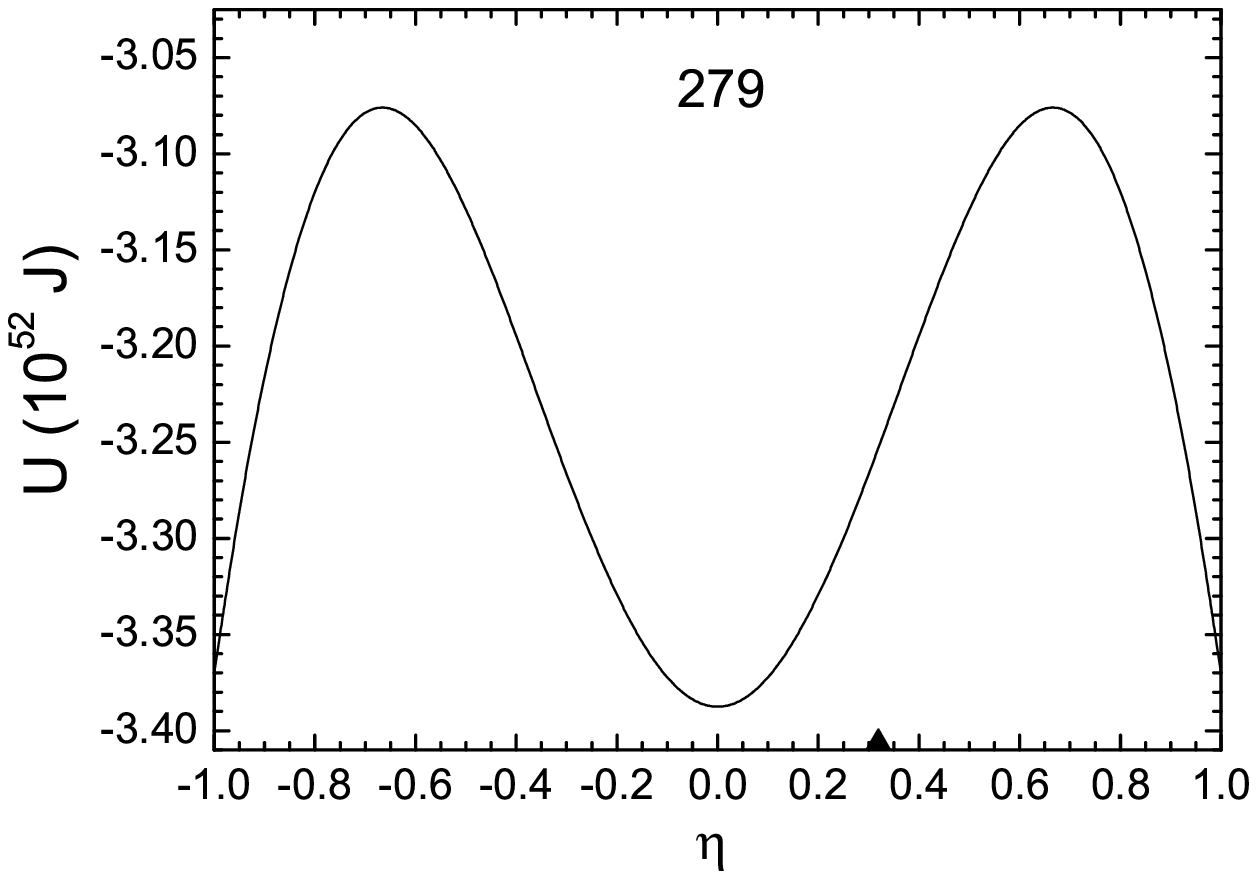}}
{\includegraphics[width=0.49\linewidth]{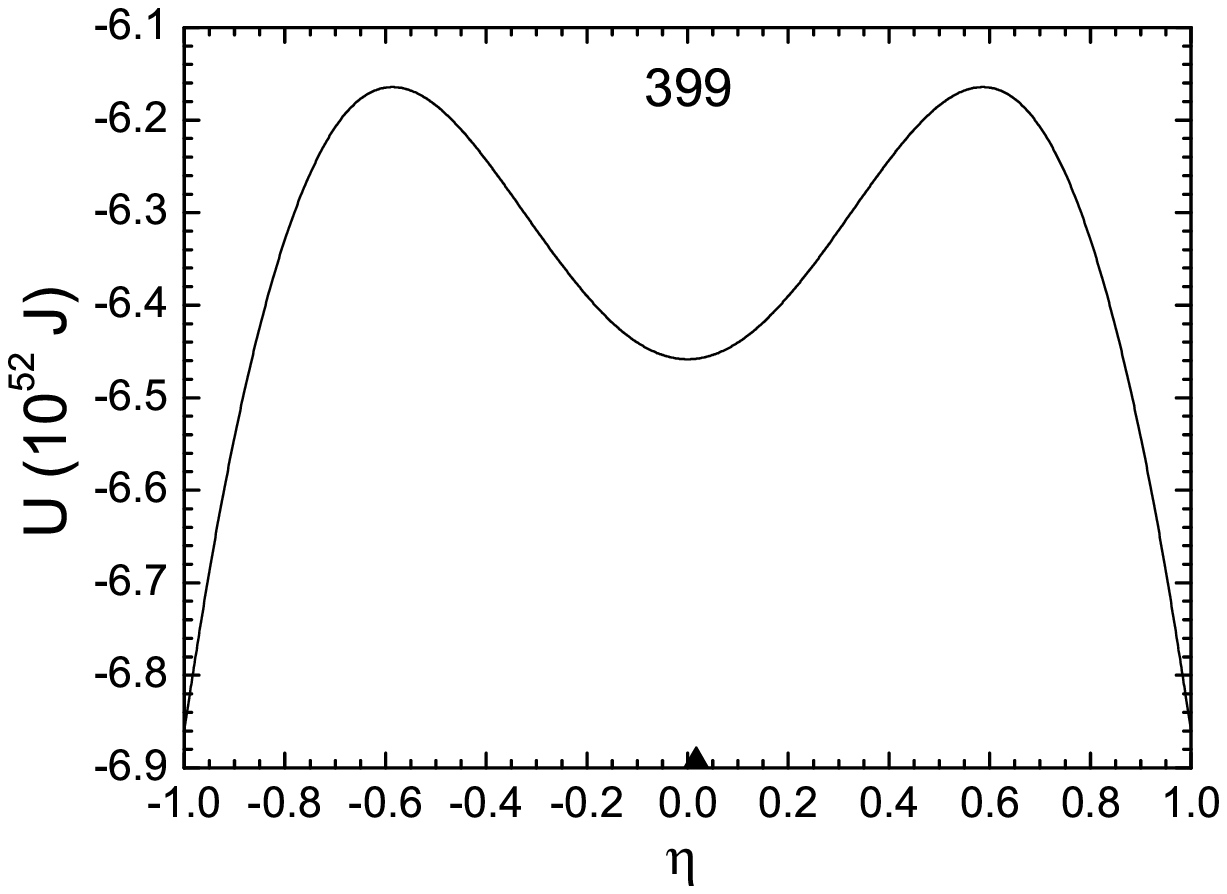}
\includegraphics[width=0.49\linewidth]{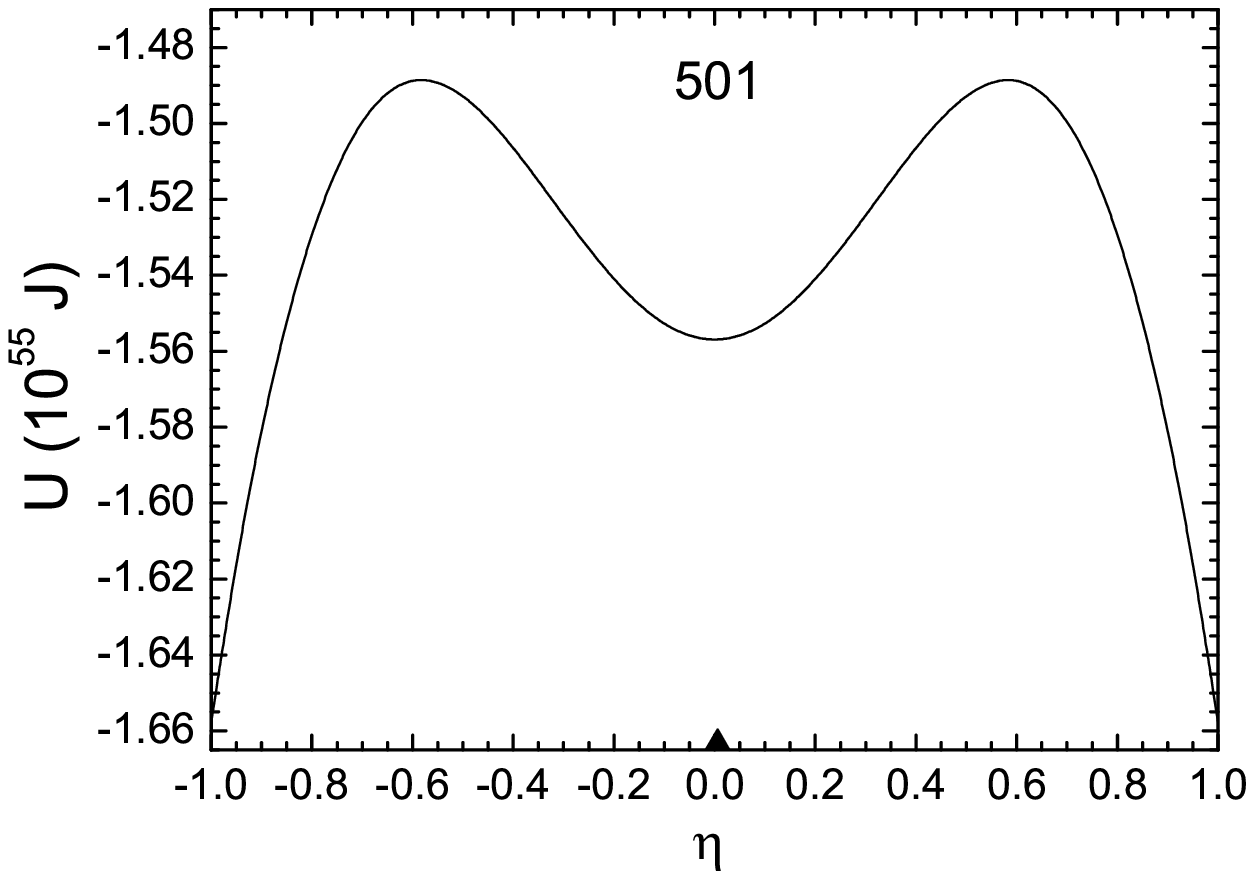}}
{\includegraphics[width=0.49\linewidth]{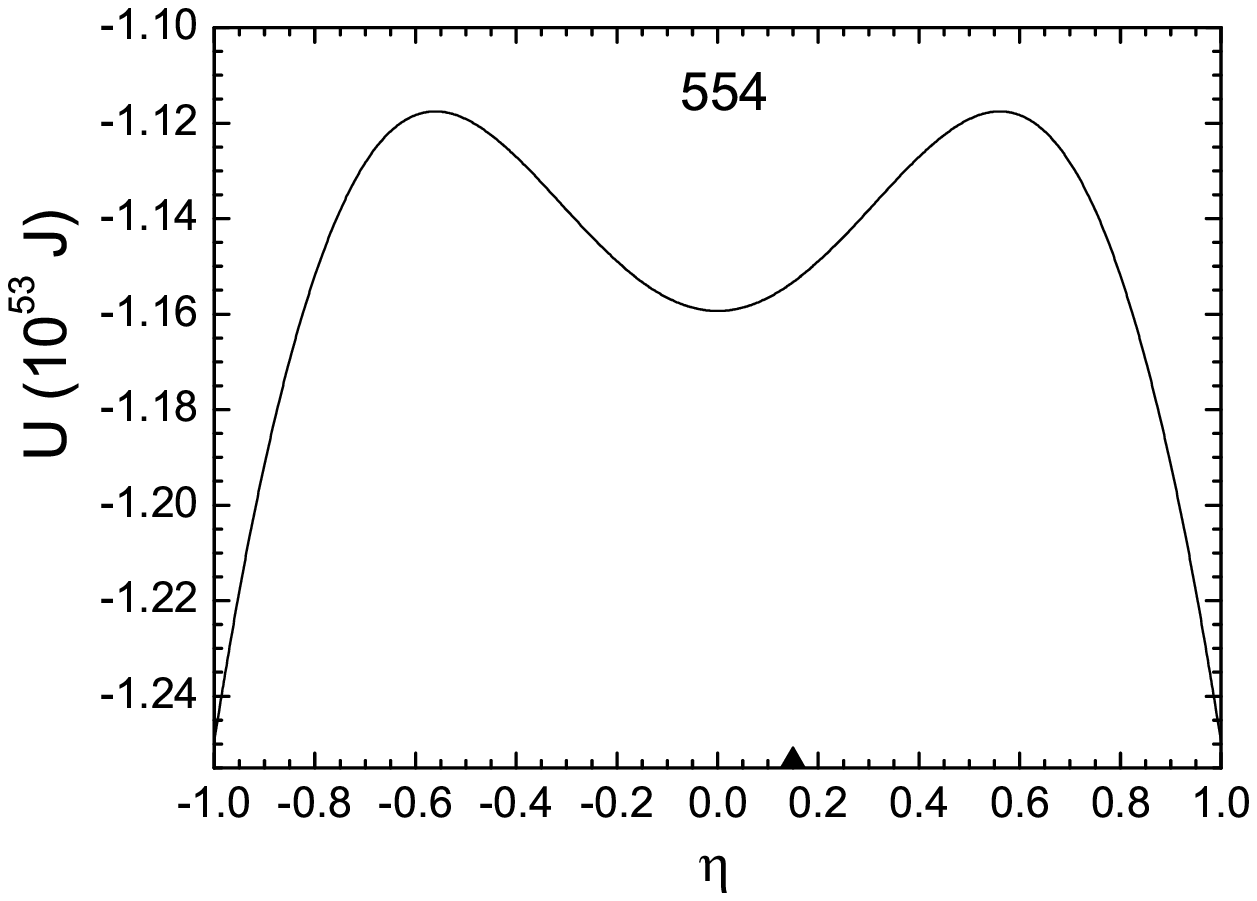}
\includegraphics[width=0.49\linewidth]{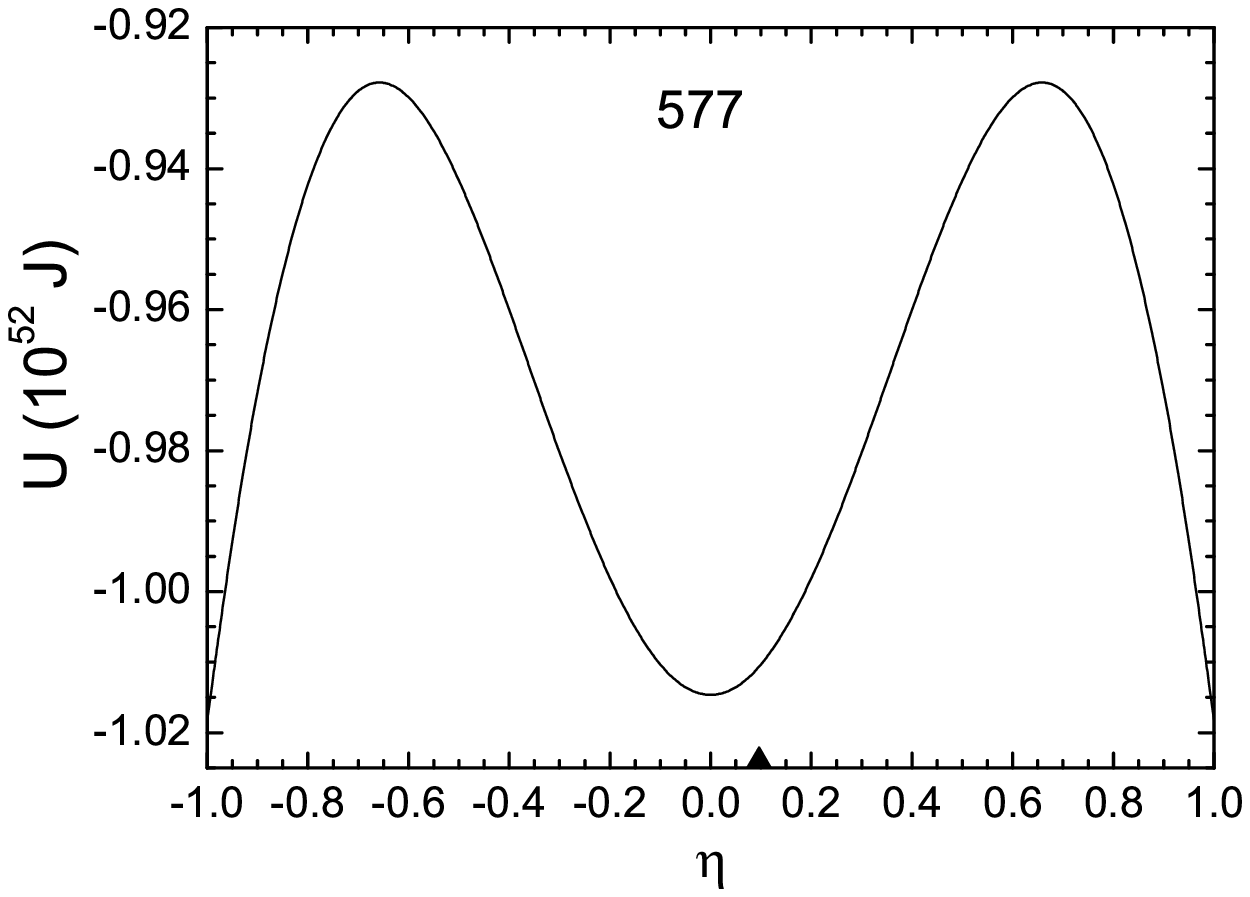}}
\caption{The same as in Fig. \ref{1_fig}, but
 for $n=2/3$.
}
\label{2_fig}
\end{figure}

\begin{figure} [ht]
\centering
{\includegraphics[width=0.49\linewidth]{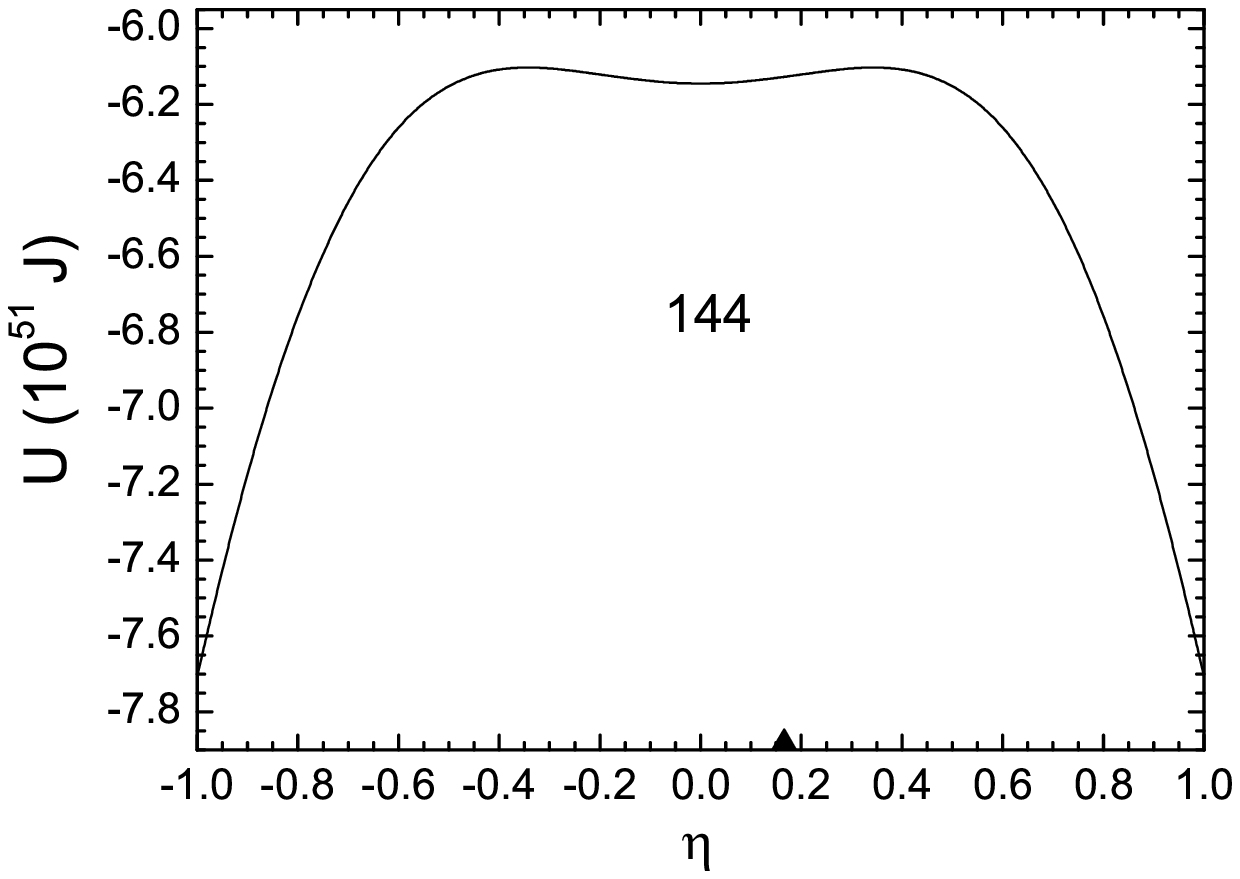}
\includegraphics[width=0.49\linewidth]{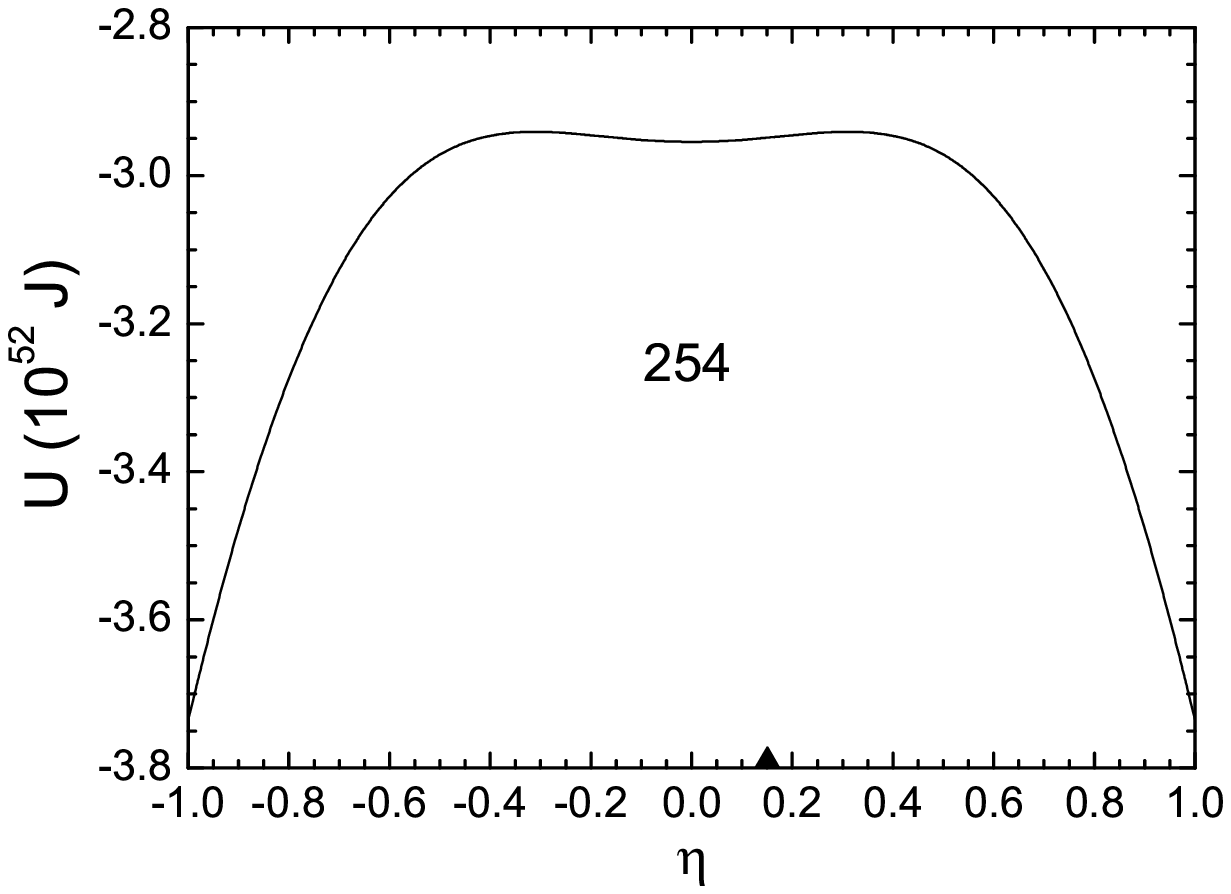}}
{\includegraphics[width=0.49\linewidth]{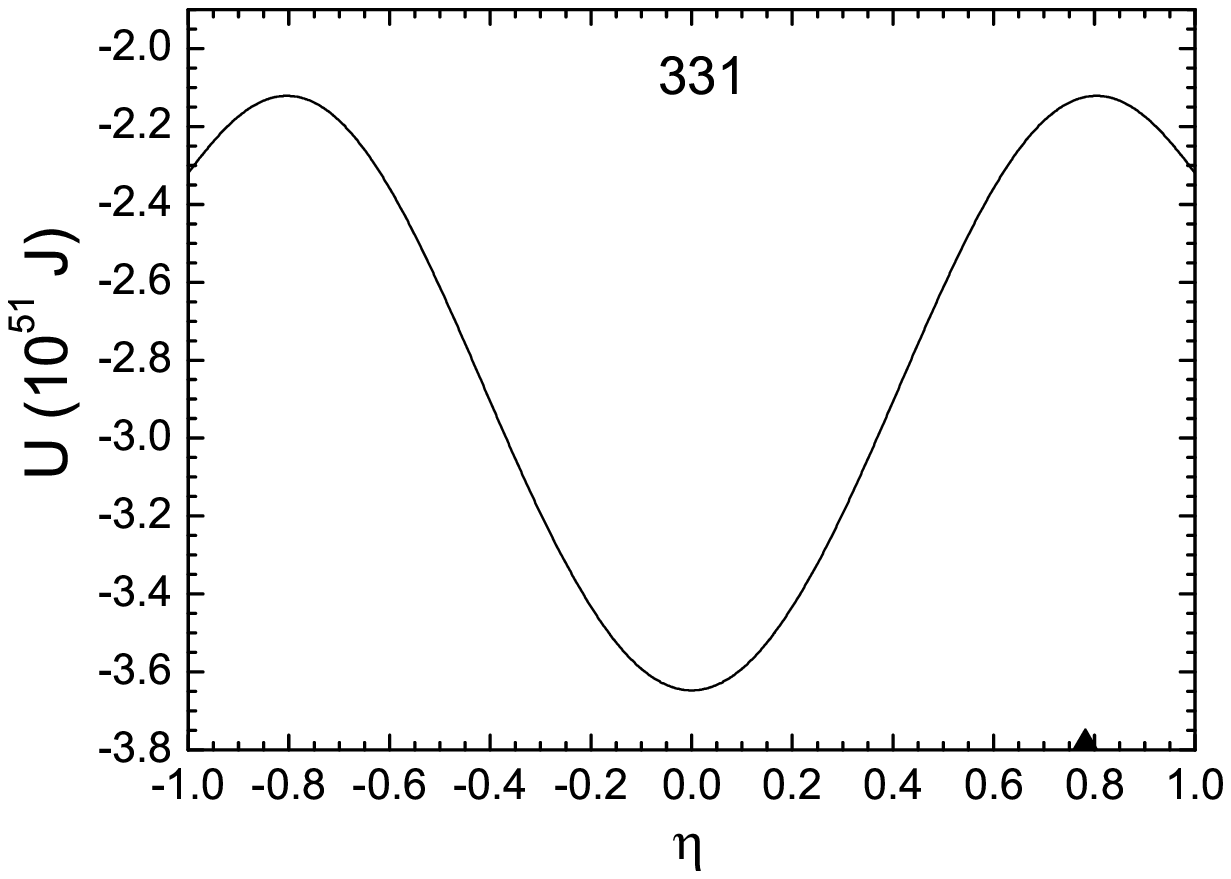}
\includegraphics[width=0.49\linewidth]{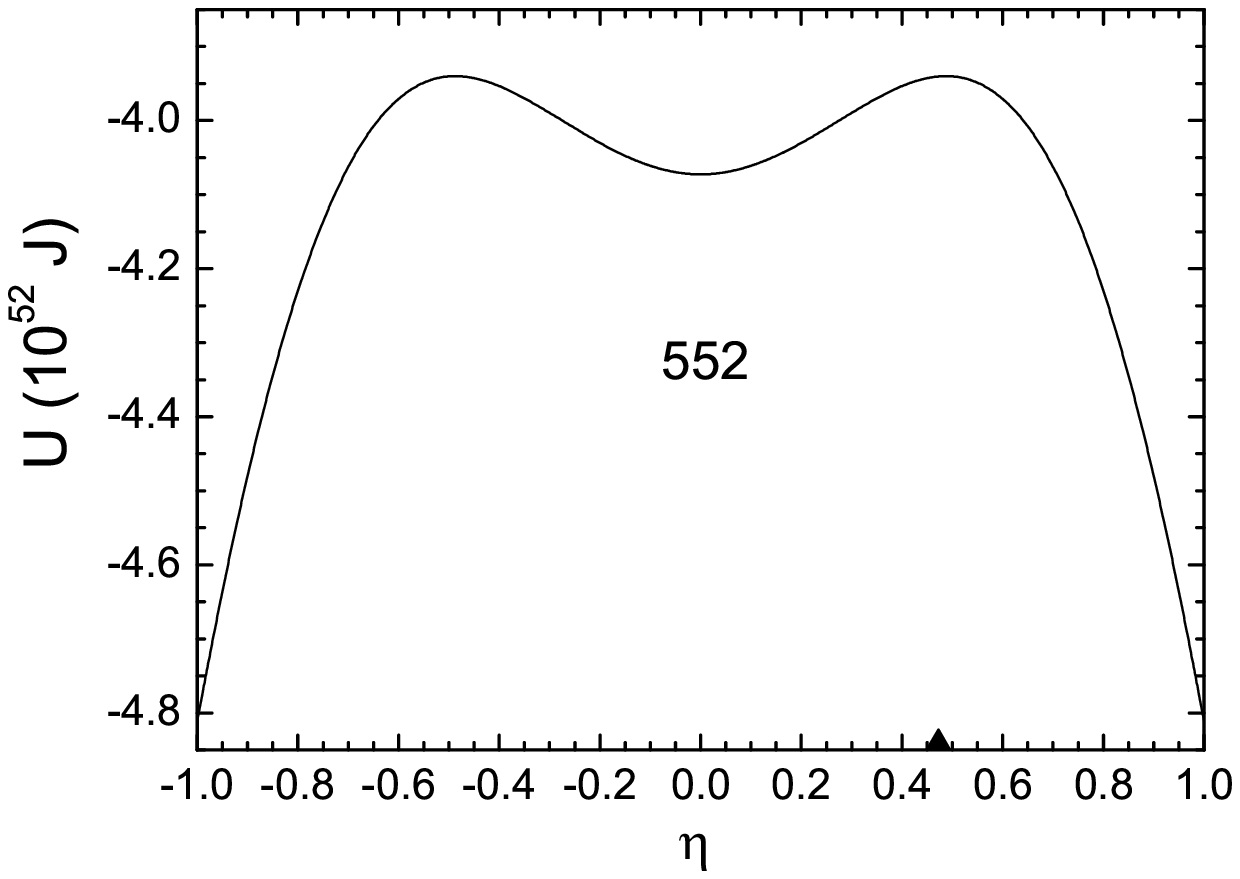}}
\caption{The same as in Fig. \ref{1_fig}, but
 for the   close binary elliptic-spiral galaxies.
}
\label{3_fig}
\end{figure}

\begin{figure} [ht]
\centering
{\includegraphics[width=0.49\linewidth]{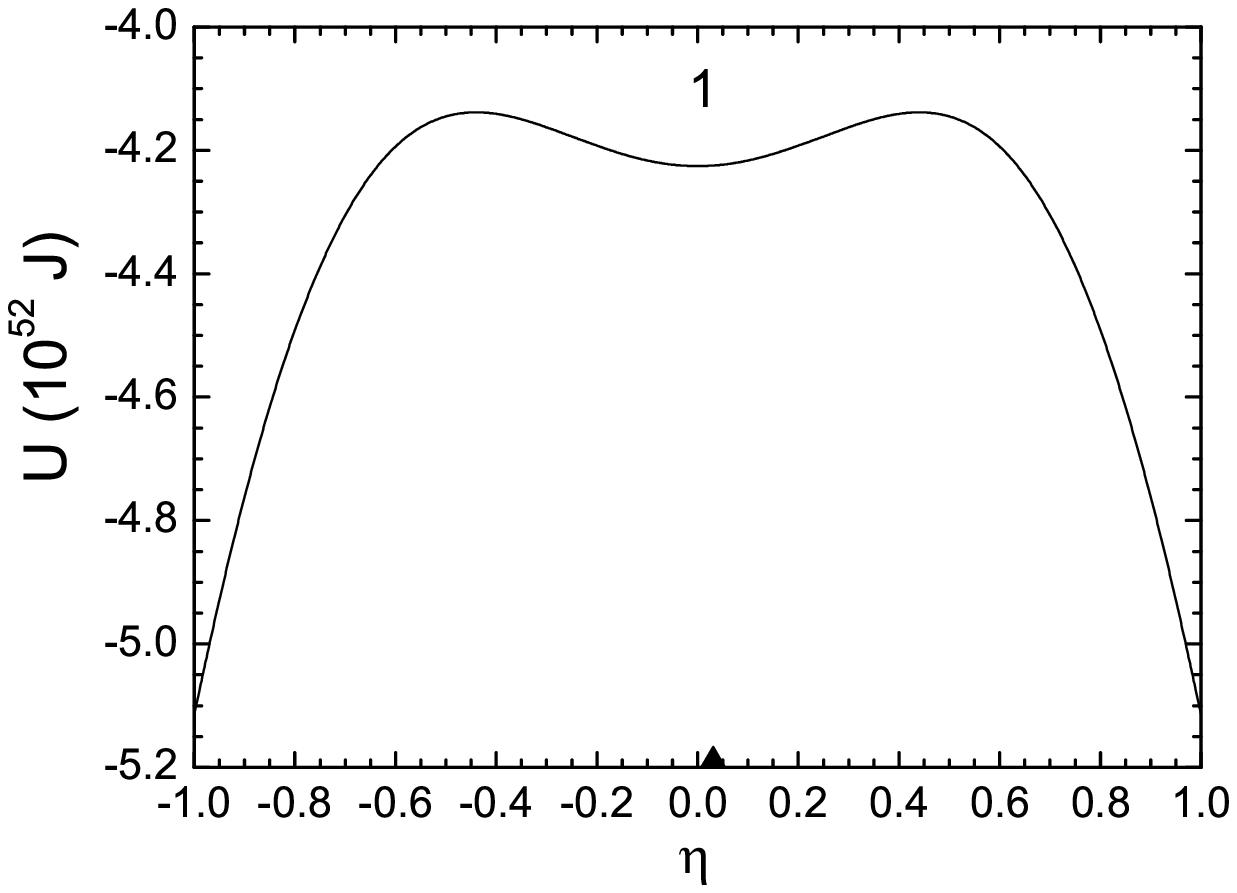}
\includegraphics[width=0.49\linewidth]{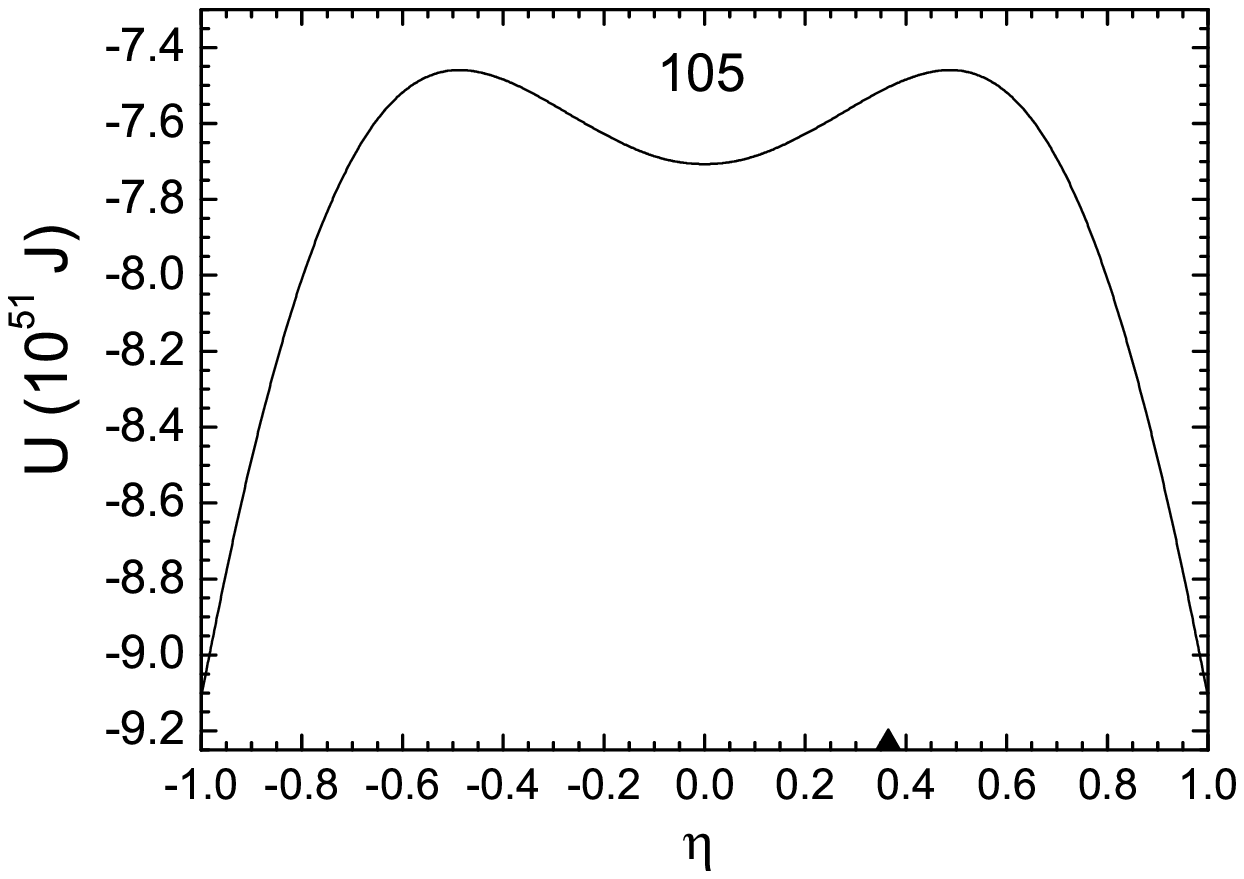}}
{\includegraphics[width=0.49\linewidth]{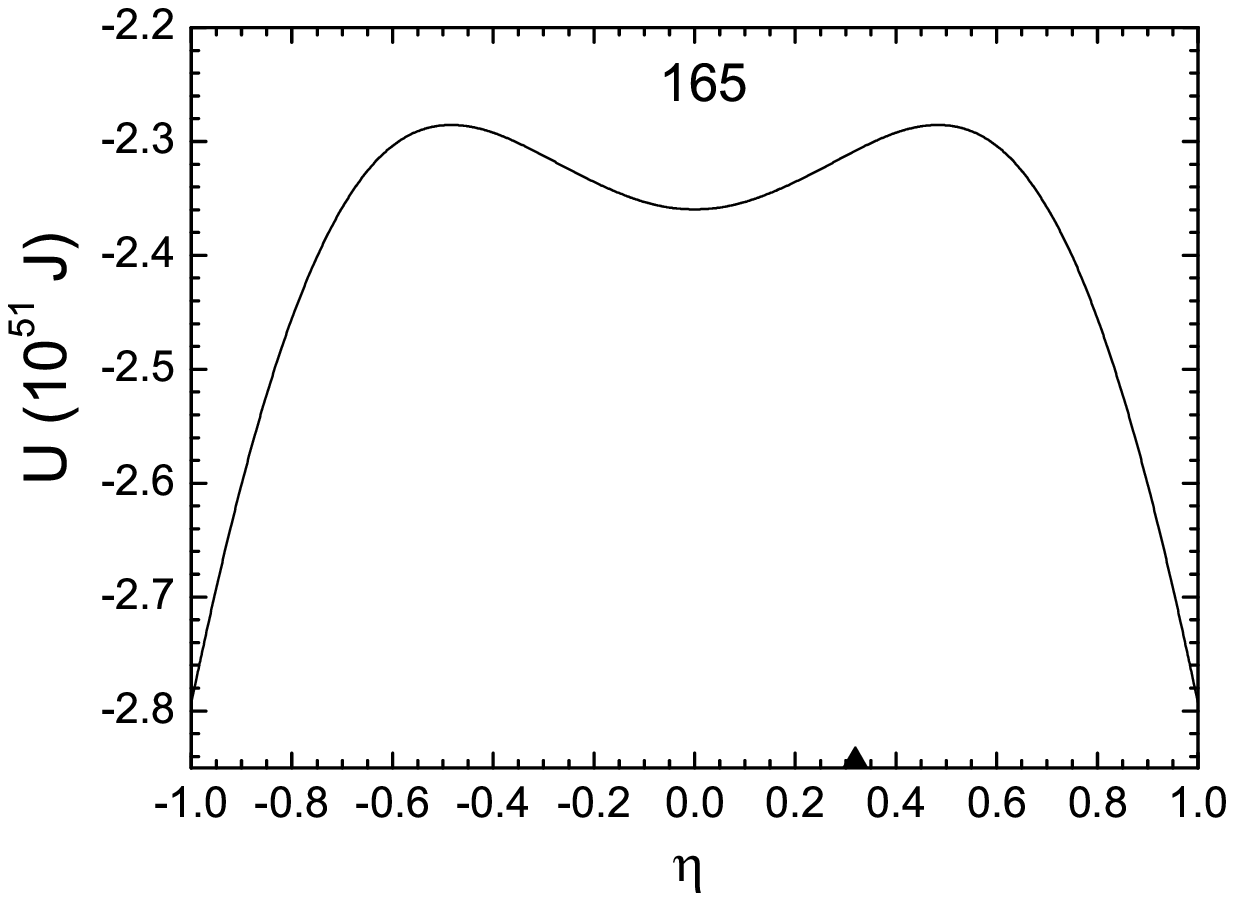}
\includegraphics[width=0.49\linewidth]{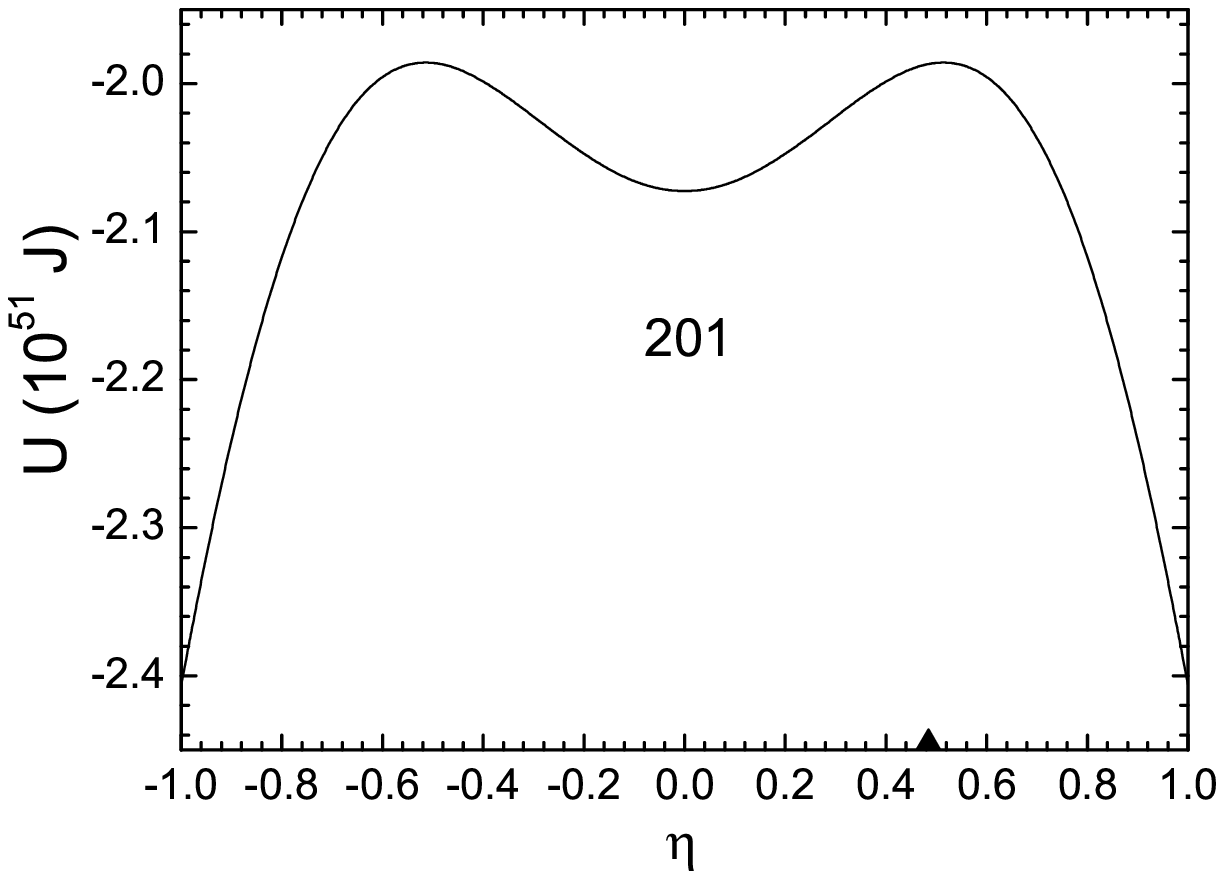}}
{\includegraphics[width=0.49\linewidth]{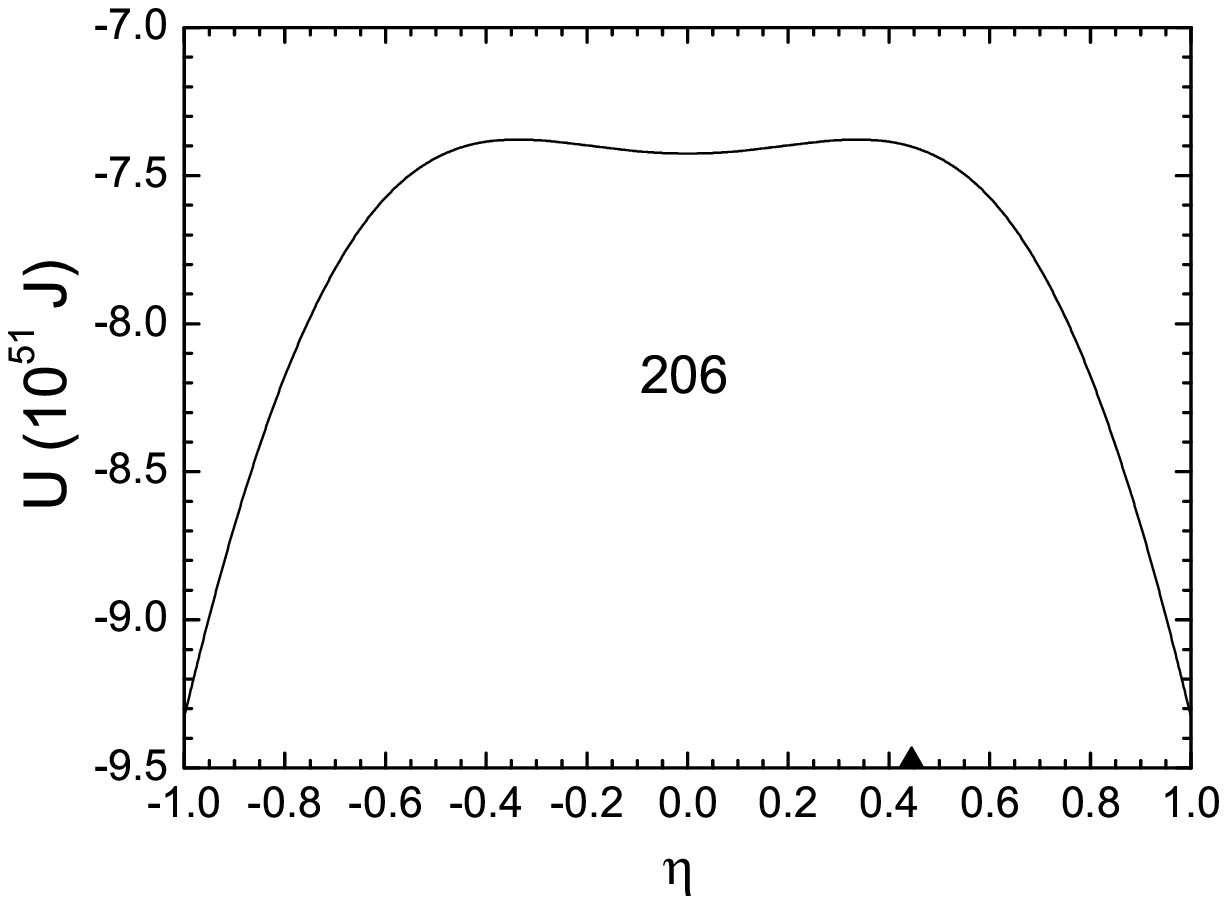}
\includegraphics[width=0.49\linewidth]{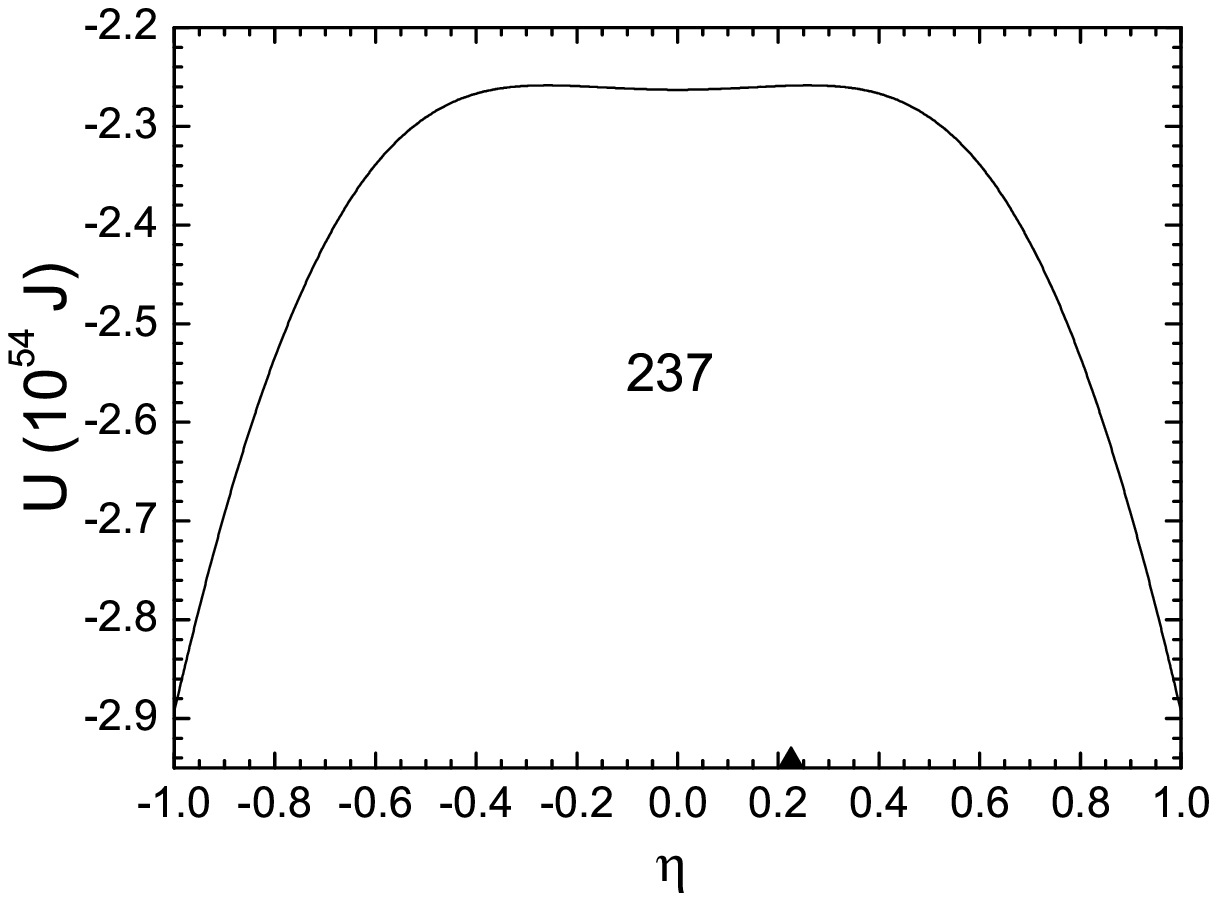}}
{\includegraphics[width=0.49\linewidth]{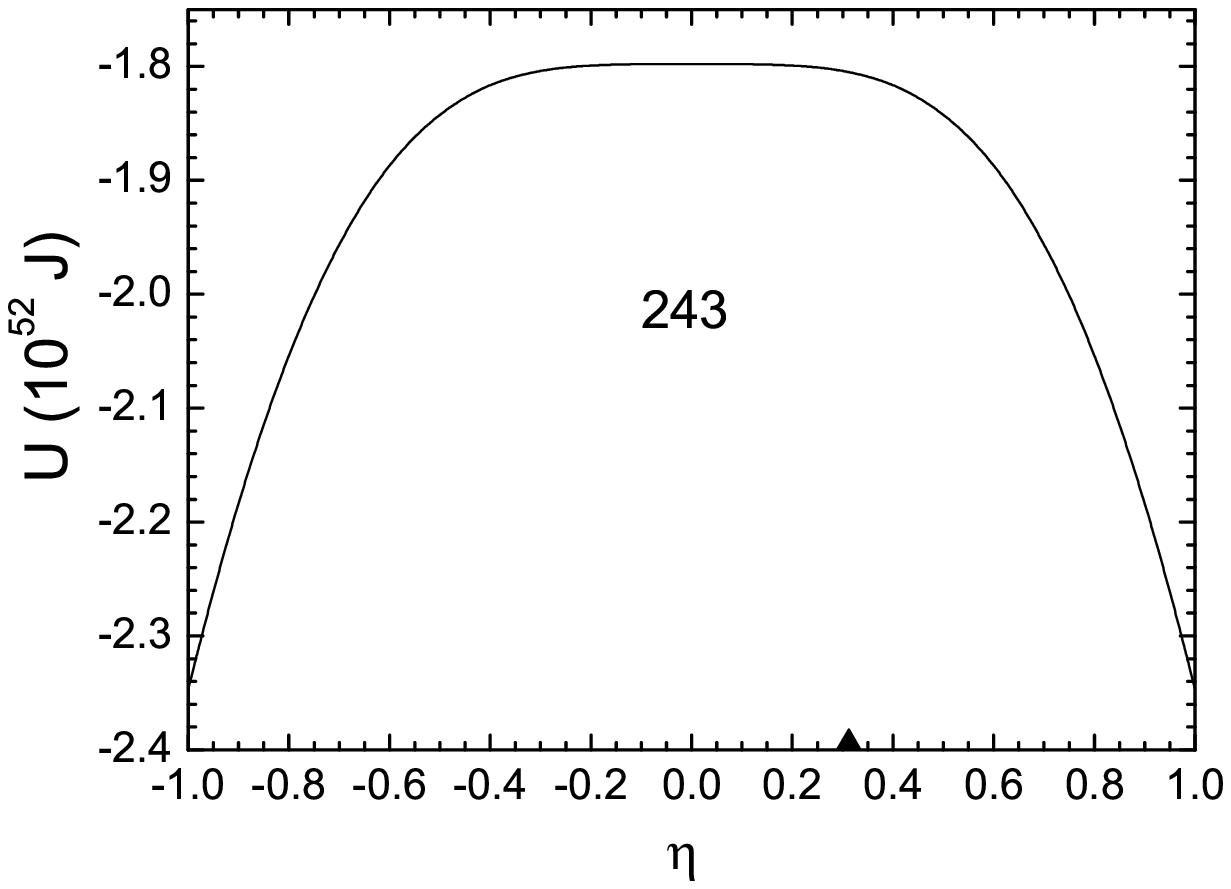}
\includegraphics[width=0.49\linewidth]{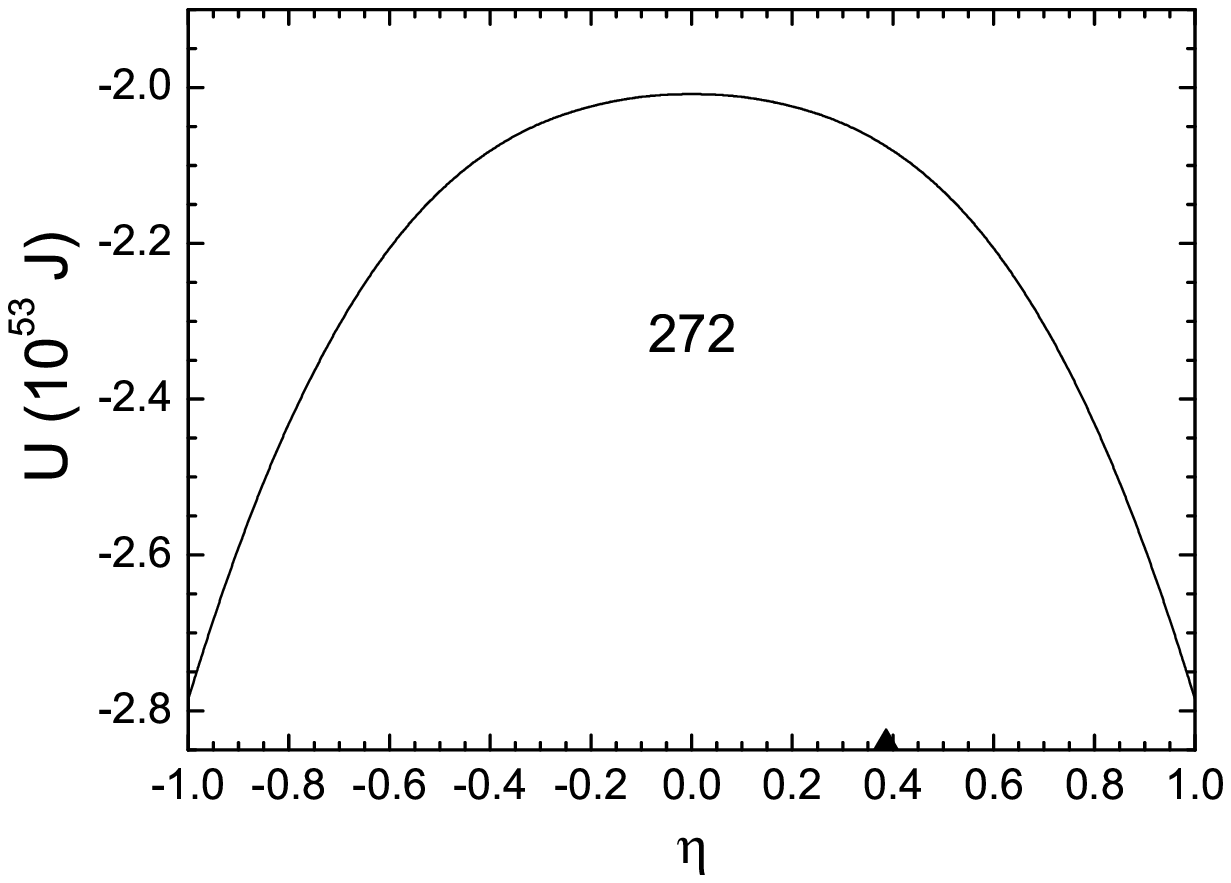}}
\caption{The same as in Fig. \ref{1_fig}, but
 for the  close    binary spiral  galaxies.
}
\label{4_fig}
\end{figure}

\begin{figure} [ht]
\centering
{\includegraphics[width=0.49\linewidth]{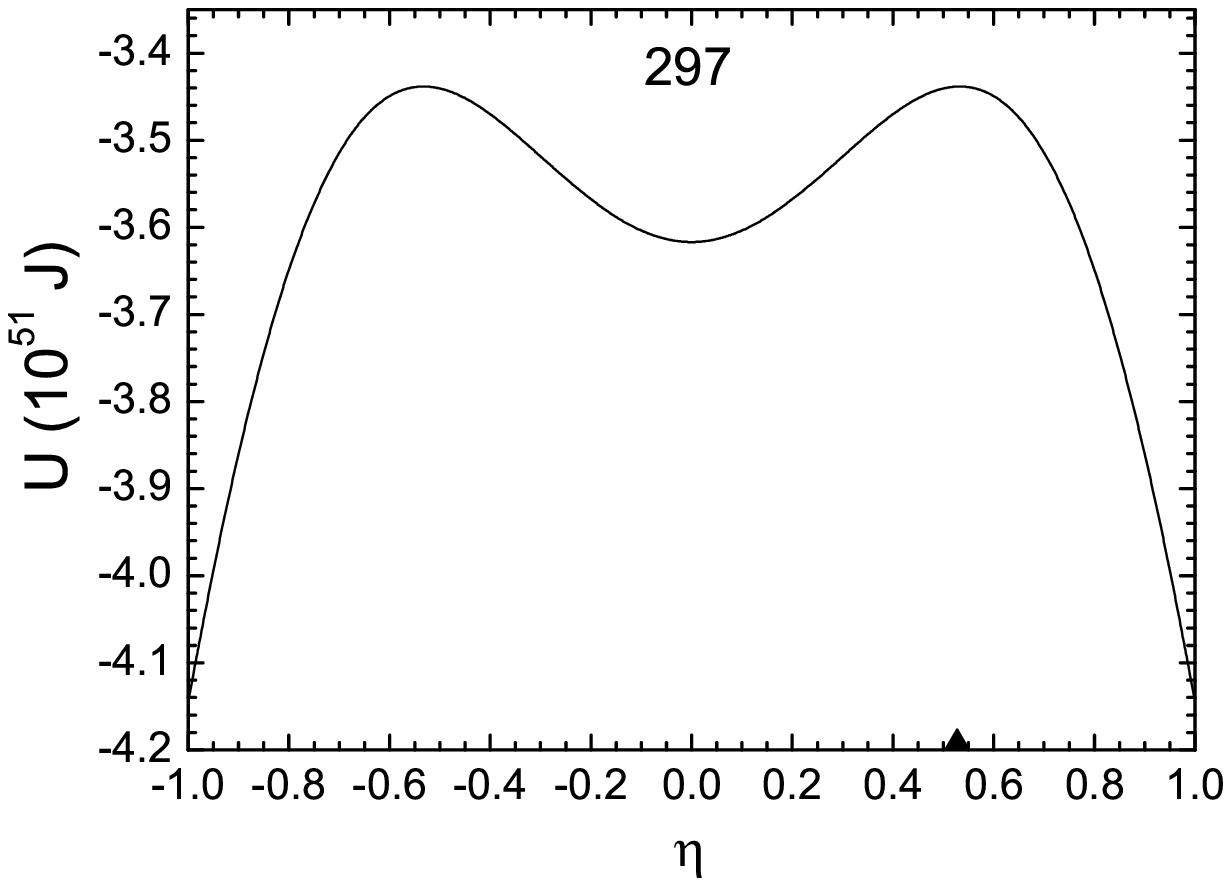}
\includegraphics[width=0.49\linewidth]{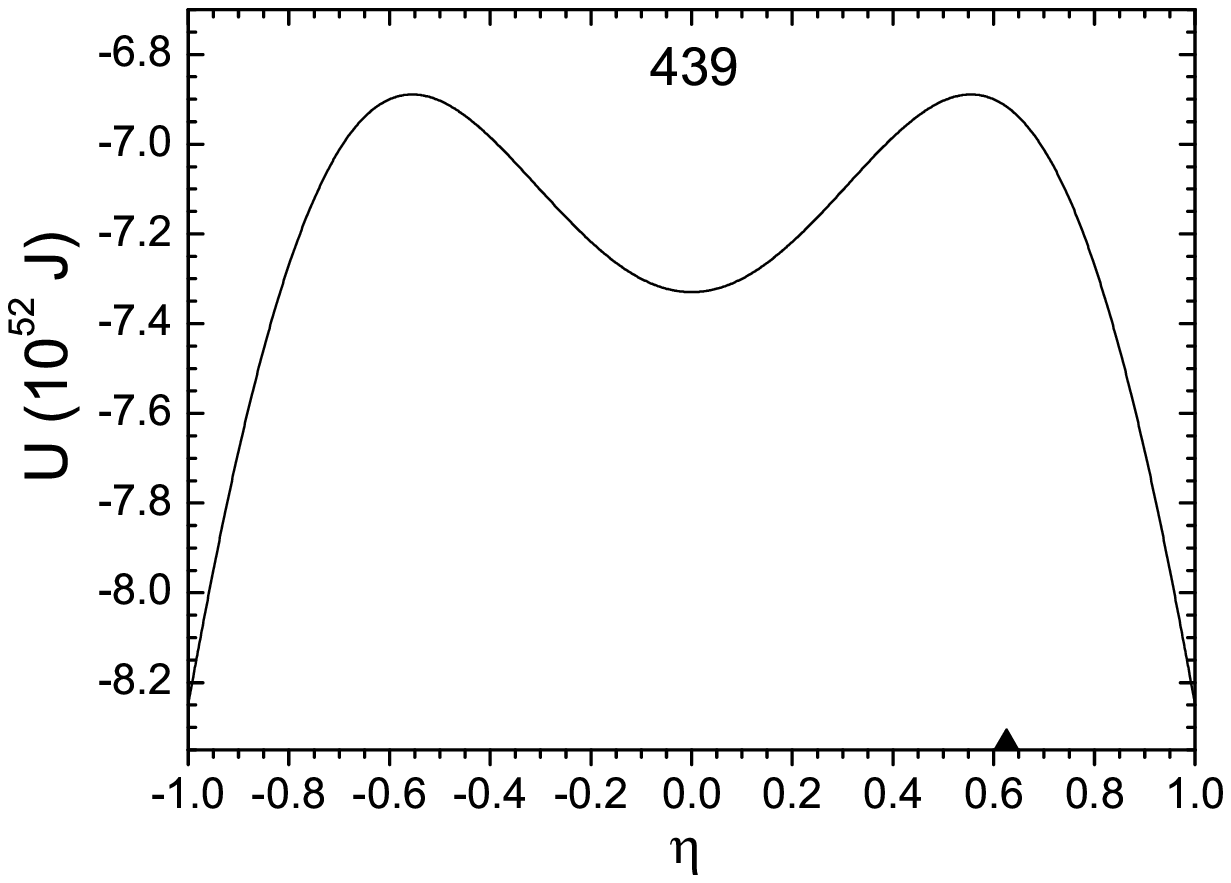}}
{\includegraphics[width=0.49\linewidth]{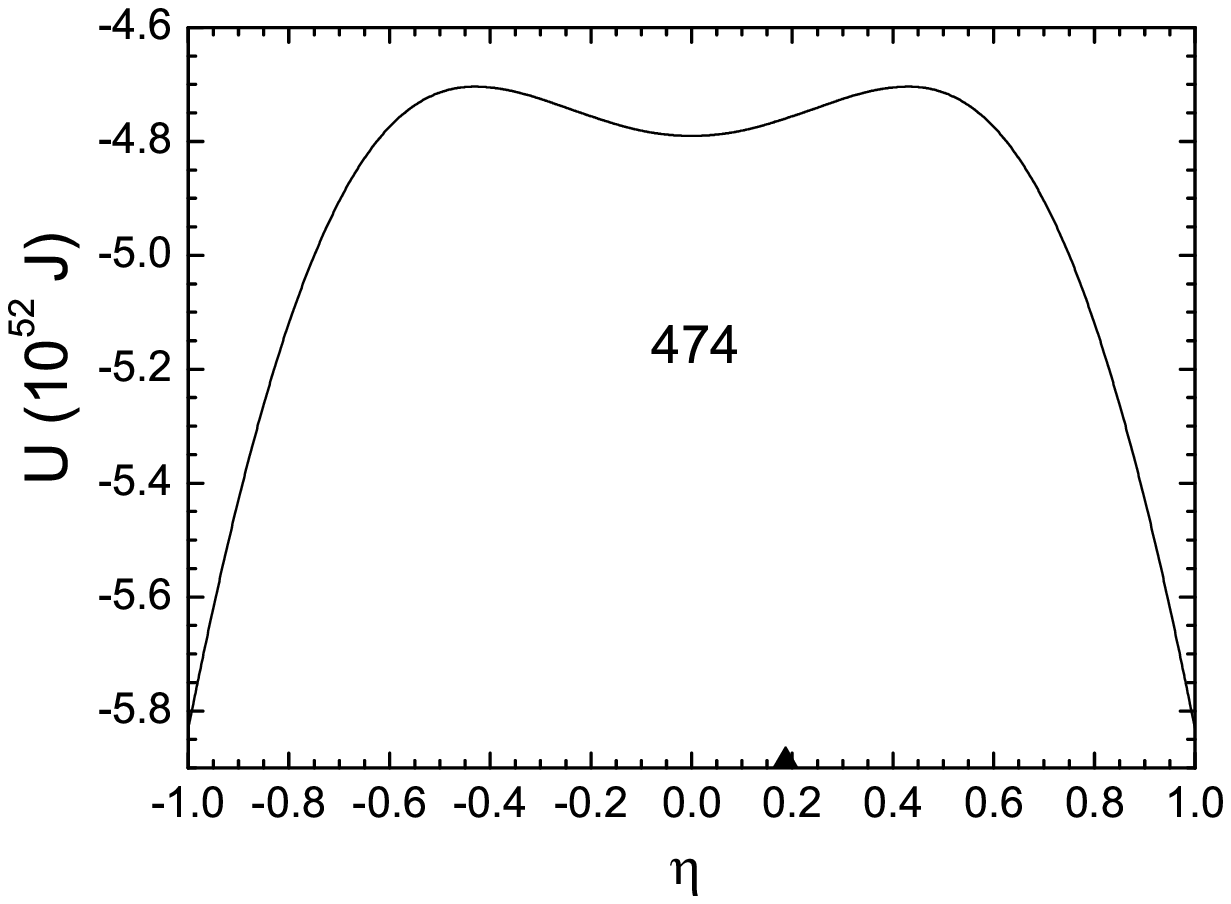}
\includegraphics[width=0.49\linewidth]{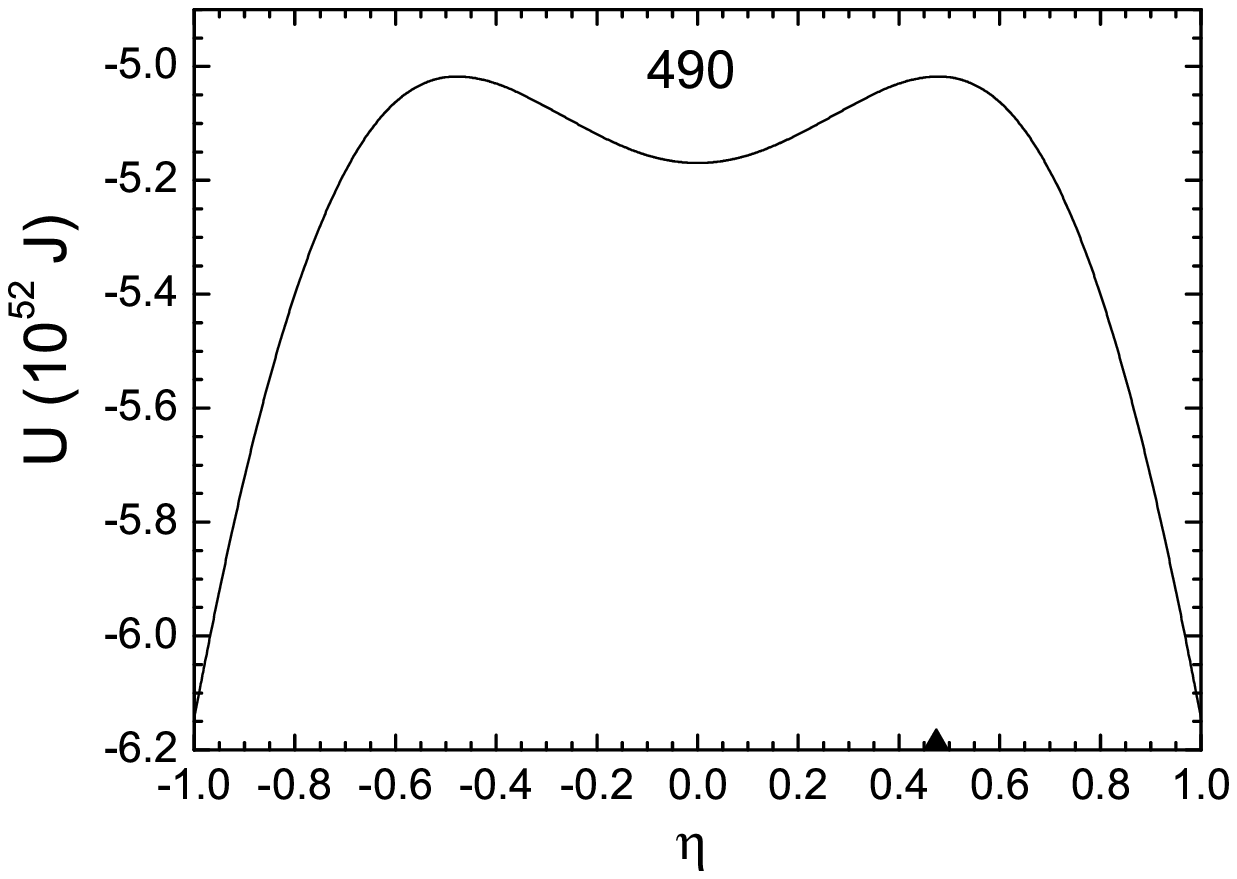}}
{\includegraphics[width=0.49\linewidth]{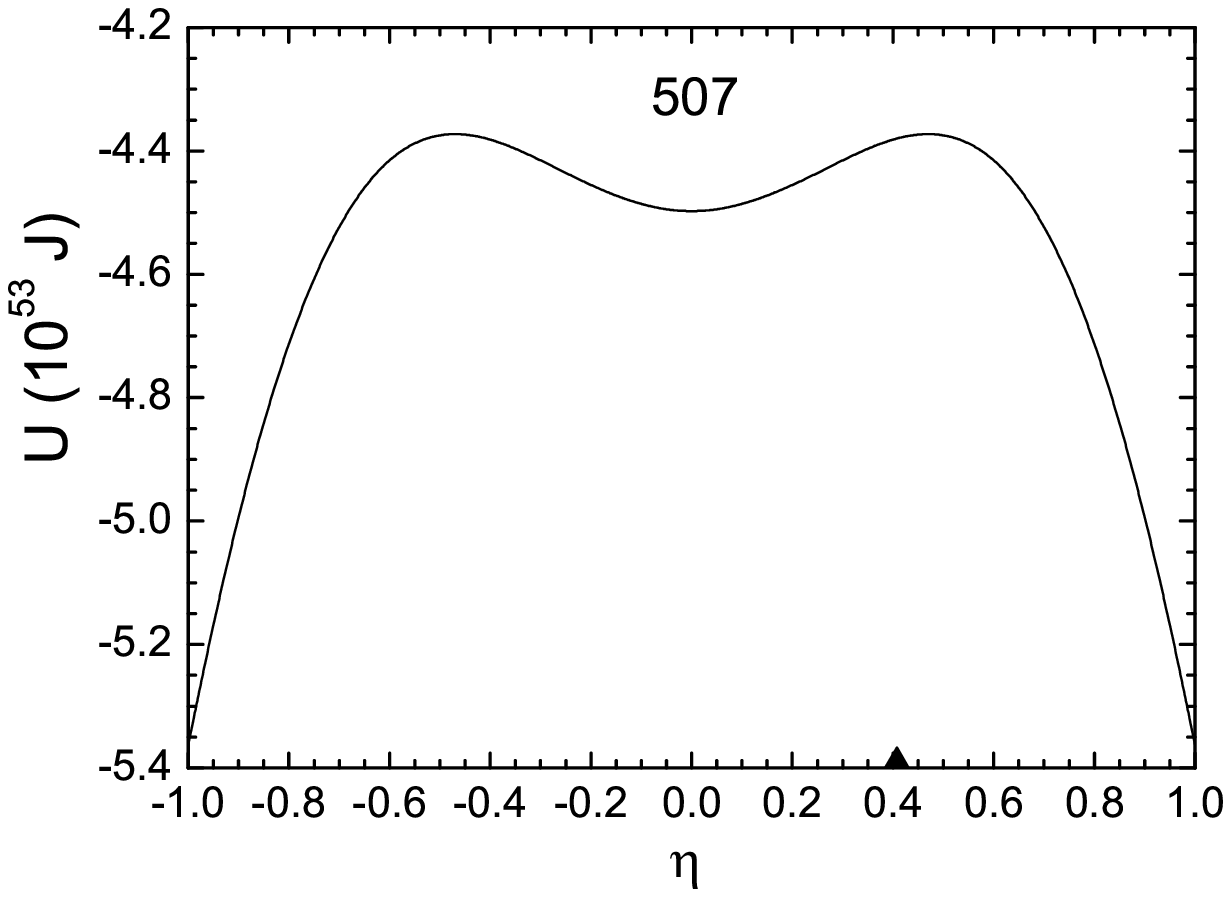}
\includegraphics[width=0.49\linewidth]{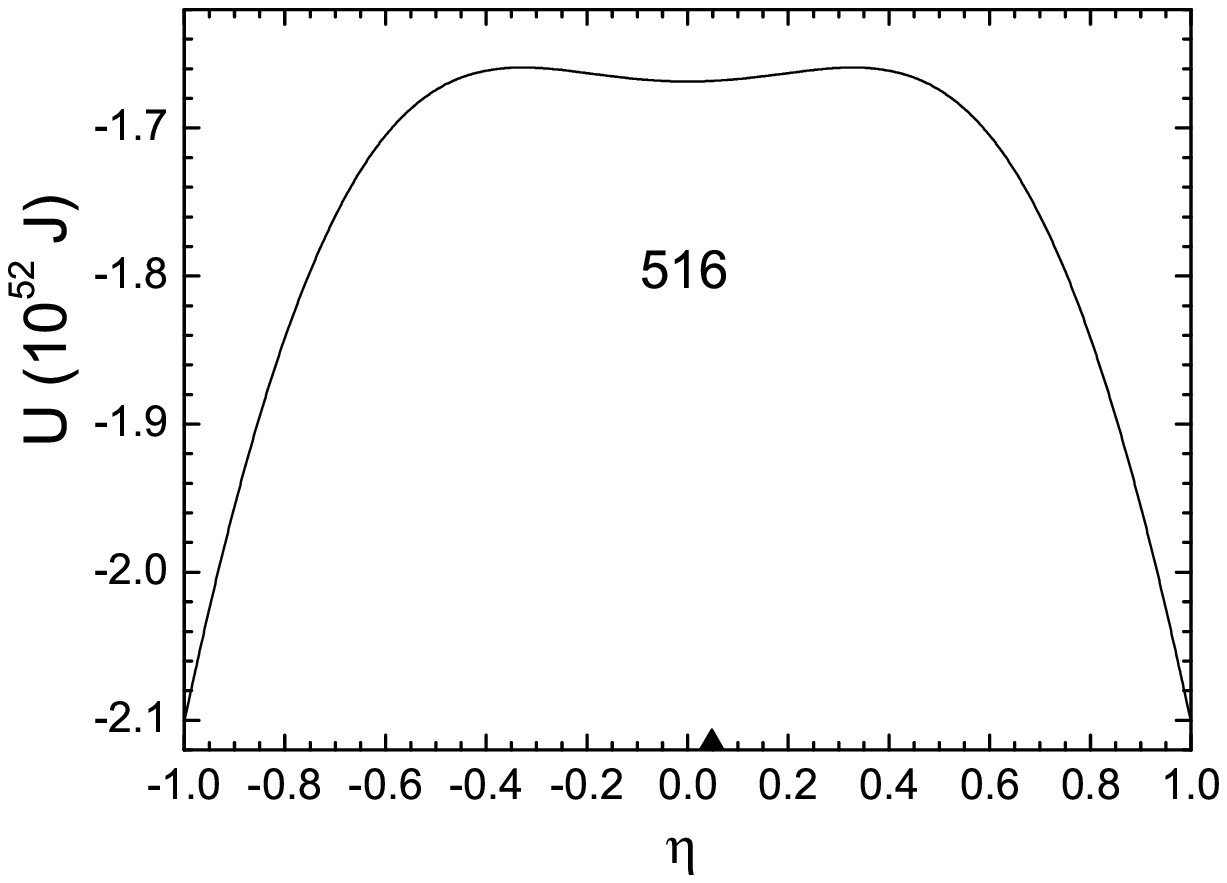}}
{\includegraphics[width=0.49\linewidth]{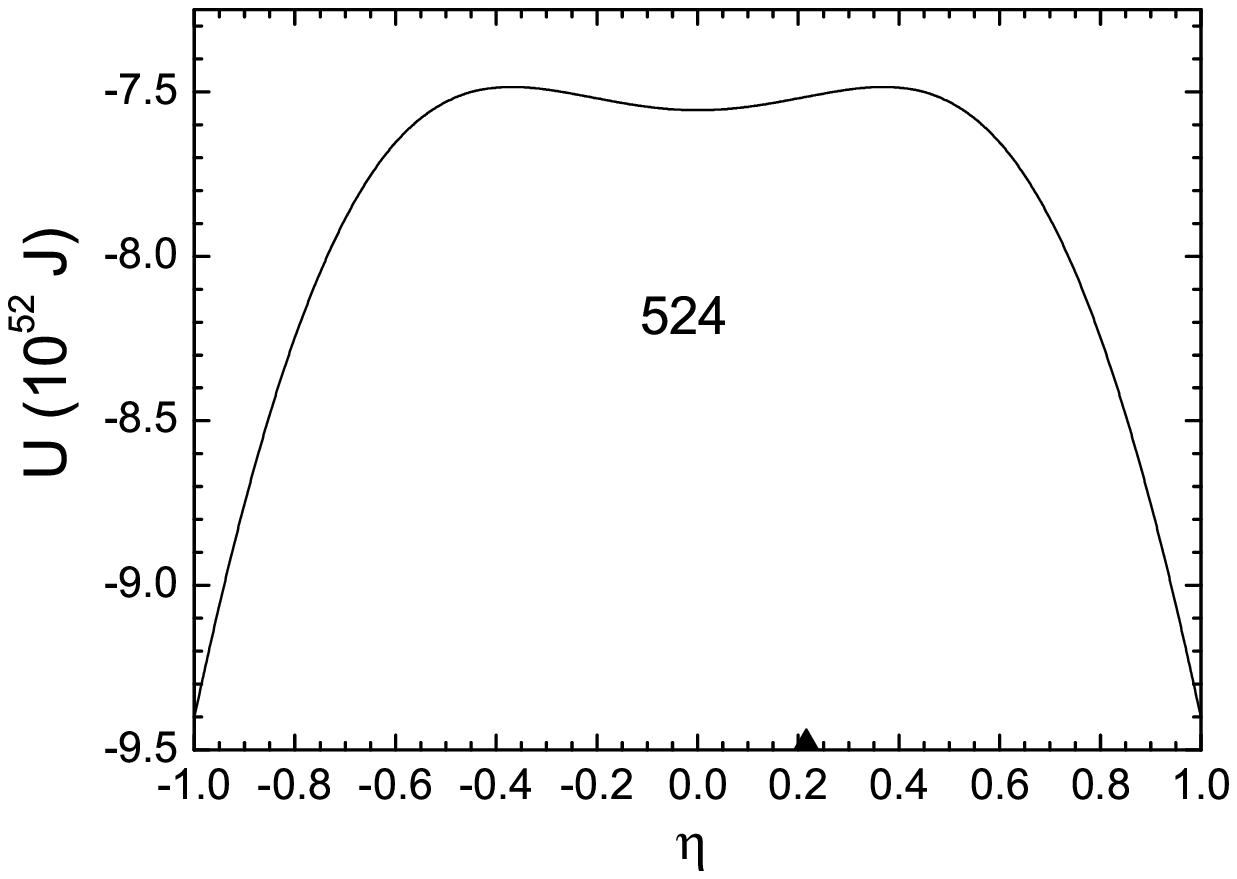}
\includegraphics[width=0.49\linewidth]{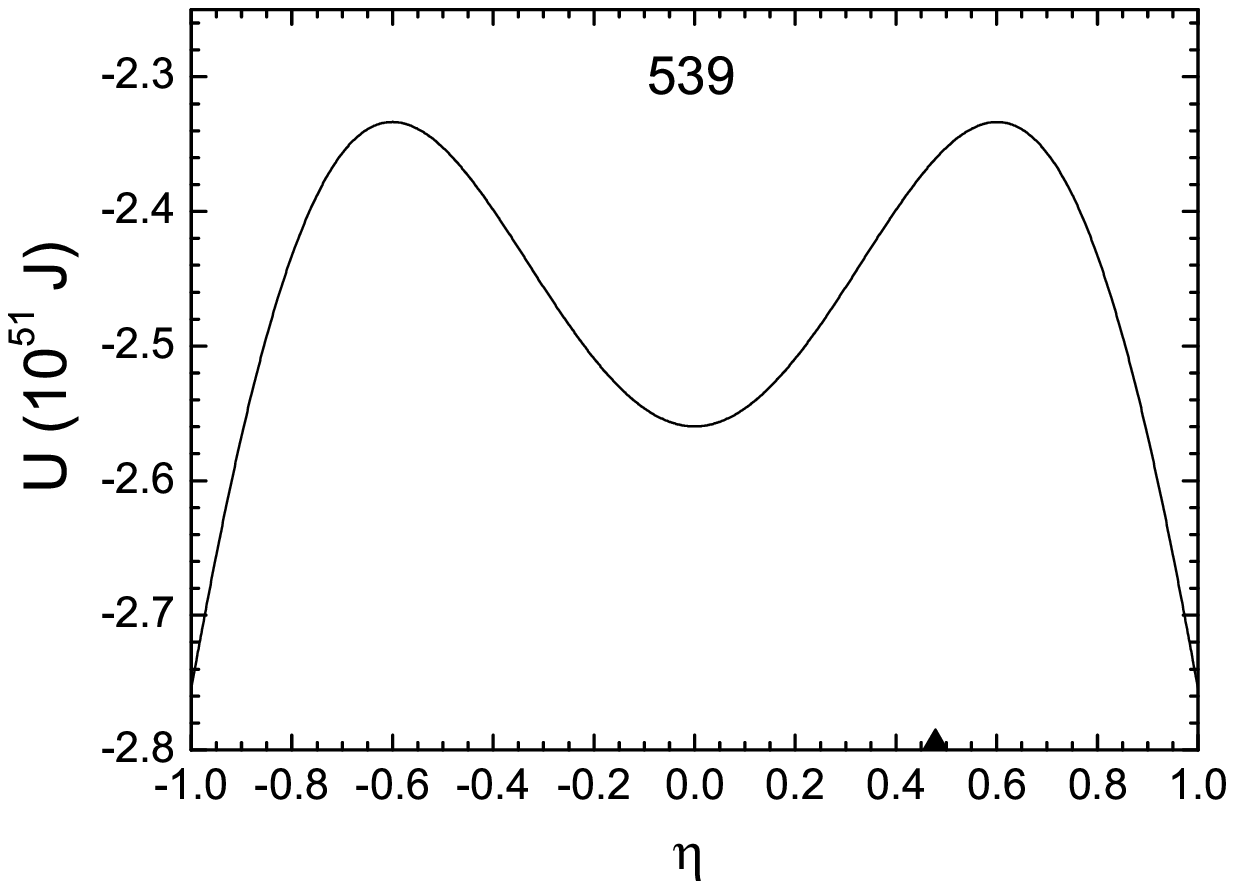}}
\caption{The same as in Fig. \ref{4_fig}, but
 for other  close    binary spiral  galaxies.
}
\label{5_fig}
\end{figure}

\newpage

\begin{table}
\small\addtolength{\tabcolsep}{0pt}
\caption{The calculated ($n=2/5$) $|\eta_i|$, $\eta_b$,    $\Delta U=U(\eta_i)-U(\eta=0)$,
 $B_\eta=U(\eta_b)-U(\eta_i)$, and observed data for
the morphological type $Ty$ of galaxy according to Hubble classification,
the total orbital mass $M/M_{\odot}$,
the projection of the linear distance $X$  between  components,
the   linear diameters $A_{25}(1i)=2R_{1i}$ and $A_{25}(2i)=2R_{2i}$ of components
of the close binary elliptic galaxies indicated by the serial numbers
in the catalogue of isolated pairs of galaxies \protect\cite{Karachentsev}.
}
\begin{tabular*}{\textwidth}{@{\extracolsep{\fill}}|c|c|c|c|c|c|c|c|c|c|}
\hline
 Di-galaxy &  $Ty$ & $\frac{M}{M_{\odot}}$ & $X$ & $A_{25}(1i)$  &  $A_{25}(2i)$  & $|\eta_i|$ & $\eta_b$ & $\Delta U$         &   $B_\eta$        \\
          &         &               &   (kpc)  &    (kpc) &     (kpc)  &            &          &    ($J$)            &         ($J$)          \\
\hline
%-------------------------------------------------------------------
194      & E-E  & $2.72\times 10^{11}$ &   27.0 & 36.9    & 29.3  & 0.28  & 0.42 & $2\times 10^{50}$      &   $2\times 10^{50}$                \\
279      & E-E  & $3.76\times 10^{11}$  &  20.6 & 27.2    & 17.5  &  0.50 & 0.59 & $3\times 10^{51}$      &   $2\times 10^{50}$                  \\
399      & E-E  & $5.84\times 10^{11}$ &   28.2 & 27.1   &  26.5  & 0.03  & 0.27 & $3\times 10^{48}$      &   $2\times 10^{50}$                   \\
501      & E-E  & $1.05\times 10^{13}$ &   38.3 & 36.0   &  35.7  & 0.01  & 0.25 & $1\times 10^{50}$      &   $3\times 10^{52}$                 \\
554      & E-E  & $1.05\times 10^{12}$  &  57.6 & 52.2   &  42.7  & 0.25  & 0.26 & $2\times 10^{50}$      &   $2\times 10^{48}$           \\
577      & E-E  & $2.27\times 10^{11}$  &  21.3 & 29.0   &  25.5  &  0.16 & 0.46 & $6\times 10^{49}$      &   $2\times 10^{50}$                      \\

\hline
\end{tabular*}
\label{tab1}
\end{table}

\begin{table}
\small\addtolength{\tabcolsep}{0pt}
\caption{The same as in Table I, but for other close binary elliptic-spiral galaxies.
}
\begin{tabular*}{\textwidth}{@{\extracolsep{\fill}}|c|c|c|c|c|c|c|c|c|c|}
\hline
 Di-galaxy &  $Ty$ & $\frac{M}{M_{\odot}}$ & $X$ & $A_{25}(1i)$  &  $A_{25}(2i)$  & $|\eta_i|$ & $\eta_b$ & $\Delta U$         &   $B_\eta$        \\
          &         &               &   (kpc)  &    (kpc) &     (kpc)  &            &          &    ($J$)            &         ($J$)          \\
\hline
%-------------------------------------------------------------------
144      & E-Sa  & $1.43\times 10^{11}$ &   17.1 & 18.3    & 16.0  & 0.17  & 0.35 & $3\times 10^{49}$      &   $3\times 10^{49}$                \\
254      & E-Sb  & $5.16\times 10^{11}$  &  47.7 & 48.9    & 43.3  &  0.15 & 0.32 & $6\times 10^{49}$      &   $8\times 10^{49}$                  \\
331      & SO-E & $8.15\times 10^{10}$ &   17.4 & 23.4   &  10.1  & 0.78  & 0.80 & $2\times 10^{51}$      &   $3\times 10^{48}$                   \\
552      & Sa-E & $5.01\times 10^{11}$ &   38.8 & 39.5   &  26.2  & 0.47  & 0.49 & $3\times 10^{50}$      &   $6\times 10^{48}$                 \\

\hline
\end{tabular*}
\label{tab1}
\end{table}

\begin{table}
\small\addtolength{\tabcolsep}{0pt}
\caption{The same as in Table I, but for close binary spiral galaxies. The values of $\Delta U$ and  $B_\eta$ are calculated if $\eta_b > |\eta_i|$.
}
\begin{tabular*}{\textwidth}{@{\extracolsep{\fill}}|c|c|c|c|c|c|c|c|c|c|}
\hline
 Di-galaxy &  $Ty$ & $\frac{M}{M_{\odot}}$ & $X$ & $A_{25}(1i)$  &  $A_{25}(2i)$  & $|\eta_i|$ & $\eta_b$ & $\Delta U$         &   $B_\eta$        \\
          &         &               &   (kpc)  &    (kpc) &     (kpc)  &            &          &    ($J$)            &         ($J$)          \\
\hline
%-------------------------------------------------------------------
1        & Sb-Sb  & $3.70\times 10^{11}$  &   9.3 & 12.1    & 11.8  & 0.28  & 0.44 & $2\times 10^{50}$      &   $6\times 10^{50}$                \\
64       & Sc-Sc  & $6.20\times 10^{11}$  &  39.5 & 55.8    & 27.4  &  0.71 & 0.75 & $2\times 10^{52}$      &   $2\times 10^{50}$                  \\
105      & Sb-Sb  & $1.89\times 10^{11}$  &  23.5 & 28.8    & 21.2  &  0.37 & 0.49 & $2\times 10^{50}$      &   $5\times 10^{49}$                  \\
165      & Sa-Sb  & $2.23\times 10^{11}$ &   10.0 & 12.9    &  9.9  & 0.32  & 0.49 & $5\times 10^{50}$      &   $2\times 10^{50}$                   \\
201      & Sa-Sb  & $1.05\times 10^{11}$ &   32.5 & 34.8   &  22.8  & 0.48  & 0.52 & $9\times 10^{49}$      &   $1\times 10^{48}$                 \\
206      & Sb-Sb  & $1.28\times 10^{11}$  &  16.8 & 13.2   &  9.0   & 0.45  & 0.34 &                        &                       \\
237      & Sb-Sb  & $4.20\times 10^{12}$  &  46.3 & 42.9   &  35.7  &  0.23 & 0.26 & $5\times 10^{51}$      &   $4\times 10^{50}$                      \\
243      & Sb-SO  & $3.02\times 10^{11}$ &   37.6 & 28.1    & 21.7  & 0.31 & 0.07 &                          &                   \\
272      & Sc-Sc  & $1.11\times 10^{12}$  &  83.5 & 32.7    & 23.6  &  0.39 & 0.00 &                        &                      \\
297      & Sc-Sc  & $1.27\times 10^{11}$ &   29.4 & 29.9   &  18.7  & 0.53  & 0.54 & $2\times 10^{50}$      &   $1\times 10^{47}$                   \\
439      & Sa-SO  & $4.95\times 10^{11}$ &   29.4 & 23.4   &  13.0  & 0.63  & 0.56 &                         &                    \\
474      & Sa-Sb  & $4.77\times 10^{11}$  &  21.7 & 27.1   &  23.3  & 0.19  & 0.43 & $3\times 10^{50}$      &   $6\times 10^{50}$           \\
490      & Sa-Sa  & $4.90\times 10^{11}$  &  30.0 & 29.6   &  19.6  &  0.47 & 0.48 & $2\times 10^{51}$      &   $6\times 10^{47}$                      \\
507      & Sb-Sa  & $1.60\times 10^{12}$ &   32.2 & 35.5   &  25.1  & 0.41  & 0.47 & $1\times 10^{52}$      &   $8\times 10^{50}$                   \\
516      & Sc-Sb  & $2.59\times 10^{11}$ &   20.0 & 21.1   &  20.3  & 0.05  & 0.33 & $4\times 10^{48}$      &   $9\times 10^{49}$                 \\
524      & Sb-Sc  & $4.83\times 10^{11}$  &  15.9 & 17.4   &  14.6  & 0.22  & 0.37 & $4\times 10^{50}$      &   $3\times 10^{50}$           \\
539      & SO-Sa  & $9.68\times 10^{10}$  &  17.8 & 25.8   &  17.0  &  0.48 & 0.60 & $2\times 10^{50}$      &   $3\times 10^{49}$                      \\

\hline
\end{tabular*}
\label{tab1}
\end{table}

\end{document}